\documentclass{JHEP3} 

\JHEPspecialurl{http://www.iop.org/EJ/abstract/1475-7516/2009/11/004}

\usepackage{graphicx}

\def\a{\alpha}

\def\d{\delta}

\def\f{\phi}

\def\l{\lambda}

\def\r{\rho}

\def\g{\gamma}

\def\th{\theta}
\def\z{\zeta}
\def\s{\sigma}
\def\ra{\rightarrow}

\def\Mfunction#1{\mathop{\rm #1}\nolimits}

\def\bm#1{\mbox{\boldmath{$#1$}}}

\title{Statistics and geometry of cosmic voids}
\author{Jos\'e Gaite\\
Instituto de Microgravedad IDR, 
ETS Ingenieros Aeron\'auticos, 
Universidad Polit\'ecnica de Madrid,
E-28040 Madrid, Spain; \email{jose.gaite@upm.es}
}

\preprint{November 4, 2009}
\received{August 26, 2009} 		
\accepted{September 30, 2009}	

\abstract{ We introduce new statistical methods for the study of cosmic voids,
focusing on the statistics of largest size voids. We distinguish three
different types of distributions of voids, namely, Poisson-like,
lognormal-like and Pareto-like distributions. The last two distributions are
connected with two types of fractal geometry of the matter
distribution. Scaling voids with Pareto distribution appear in fractal
distributions with box-counting dimension smaller than three (its maximum
value), whereas the lognormal void distribution corresponds to multifractals
with box-counting dimension equal to three.  Moreover, voids of the former
type persist in the continuum limit, namely, as the number density of
observable objects grows, giving rise to lacunar fractals, whereas voids of
the latter type disappear in the continuum limit, giving rise to non-lacunar
(multi)fractals.  We propose both lacunar and non-lacunar multifractal models
of the cosmic web structure of the Universe.  A non-lacunar multifractal model
is supported by current galaxy surveys as well as cosmological $N$-body
simulations.  This model suggests, in particular, that small dark matter halos
and, arguably, faint galaxies are present in cosmic voids.  }


\begin{document}

\section{Introduction}
\label{intro}

The large scale structure of the Universe is formed by matter clusters,
filaments and sheets, and also cosmic voids.  Cosmic voids are the counterpart
of matter structures.  They have arisen in observations of the galaxy
distribution and have gradually become the subject of deep theoretical
investigations.  Early studies of cosmic voids were conditioned by the small
number of galaxies surveyed then.  For example, Otto et al.\ \cite{Otto}
studied the significance of cosmic voids, to distinguish them from
fluctuations of a homogeneous galaxy distribution.  They developed
probabilistic methods for the study of voids but they actually concluded that
cosmic voids were not really significant.  Betancort-Rijo \cite{Bet} improved
and generalized their methods and reached the opposite conclusion.  The r\^ole
of voids as a basic ingredient of the large scale structure is now well
established \cite{Einasto,Kauffmann,ElAd-Pir}.

However, the definition of what constitutes a void is still imprecise.
Originally, voids were described as large regions devoid of galaxies and found
by visual inspection. As this is hardly satisfactory, an objective way of
identifying voids has been long sought.  To identify a void in a point
distribution, one needs to decide if it is to be empty or it can have a few
points inside and one further needs to determine its shape. The simplest
option, preferred by Otto et al.\ \cite{Otto} and Einasto et al.\
\cite{Einasto}, is to define voids as empty spheres. Kauffmann and Fairall
\cite{Kauffmann} also demand that voids be empty but devise a void-finder
that allows more general shapes, with the intention to find approximately
ellipsoidal voids. ElAd and Piran \cite{ElAd-Pir} devise an even more
elaborate method: they assume that there is a set of highly clustered ``wall
galaxies'' and implement a procedure to select them before the void-finding
phase, which joins empty spheres with a ``thinness'' limitation.  Therefore,
the found voids are not empty of galaxies and can have fairly complex shapes.

The preceding definitions of voids ignore that galaxy voids can contain dark
matter.  Although there is no substantial observational knowledge of the
geometry of voids in the dark matter distribution, we have information from
cosmological $N$-body simulations.  For example, Gottl\"ober et al \cite{Gott}
re-simulate voids with higher resolution and find structures inside them, in a
self-similar pattern.  This suggests that the definition of voids as empty
regions of simple shape is not appropriate in this case.  Shandarin, Sheth \&
Sahni \cite{Shan} and Sheth \& van de Weygaert \cite{Sheth-vdW} define voids
as under-dense regions of a continuous density field, such that they are
complementary to matter clusters. With this definition, voids contain matter
and have very complex shapes.  Shandarin et al.\ \cite{Shan} and Sheth \& van
de Weygaert \cite{Sheth-vdW} extend methods that have been used successfully
for the study of clusters to the study of voids. In fact, under-densities and
over-densities are symmetrical in a Gaussian random field.

Colberg et al \cite{Col_etal} have presented an overview of various void
definitions and have made a comparison of the results of applying the
corresponding void finding algorithms to the same data set.

The choice of a simple definition of voids, in particular, their definition as
empty spheres, is convenient for statistical studies of galaxy voids.  On the
other hand, the geometrical aspects of dark-matter voids that arise in
cosmological simulations are difficult to relate to the statistics of
spherical voids.  The definition of dark-matter voids as under-densities in a
Gaussian field is appealing but difficult to connect with any notion of empty
voids.  In fact, a Gaussian density field is a valid description only in the
linear regime of gravitational clustering, but empty or nearly empty voids are
very nonlinear structures.  Therefore, a suitable description of the geometry
of voids demands the use of nonlinear models of the cosmic structure.  We
focus on fractal models, for they are well founded and provide a useful
description of voids.

Fractal models of the large-scale structure of the Universe were introduced by
Mandelbrot \cite{Mandel} and have been well studied
\cite{Borga,Jones-RMP,Pietro_book}, but they are still controversial
\cite{Jones-RMP}.  The major controversy concerns the transition to
homogeneity.  Mandelbrot \cite{Mandel}, inspired by old hierarchical models of
the Universe without transition to homogeneity, has indeed suggested that
there might not be an outer cutoff to the fractal scaling range.  Assuming
that there is a scale of homogeneity, scale invariance is nevertheless
justified on smaller scales by observational and theoretical arguments.  The
observation of a power-law two-point correlation function of galaxies in a
range of scales \cite{Pee1} was one of Mandelbrot's motivations and is still a
strong argument. Furthermore, there are evidences of scaling in higher order
correlation functions \cite{Jones-RMP}.

Theoretical arguments for scale invariance are based on the absence of scales
in the gravitational dynamics of collision-less cold dark matter (CDM). This
nonlinear dynamics indeed gives rise to a hierarchical formation of structures
on every scale up to the homogeneity scale, which grows with time.  Thus, we
can reasonably expect that the resulting structure consists in a
(multi)fractal attractor of the nonlinear dynamics.  Structure formation in
CDM models is studied with $N$-body simulations, which support a multifractal
model in a range of scales \cite{Valda,Colom,Yepes,I4,I5}.  This range is
restricted by the intrinsic limitations of these $N$-body simulations (even of
state-of-the-art simulations).  The limitations are much less stringent for
one-dimensional cosmological dynamics, which is simulated by Miller et al
\cite{Miller}, finding multifractal structure in a very long range of scales.
One last but not least argument for scale invariance is that the cosmic web
produced by the adhesion model \cite{adhesion} is found to have multifractal
features \cite{V-Frisch,Bou-M-Parisi}.

Mandelbrot \cite{Mandel} has also introduced the notion of fractal holes
(``tremas'') and considered its application to the galaxy distribution.  In
particular, he has introduced the notion of fractal {\em lacunarity} as a
measure of the size of voids in fractals of equal dimension. For any given
dimension, one can construct a set of different fractals with progressively
decreasing lacunarity, leading to the possibility of a {\em non-lacunar}
fractal. Mandelbrot \cite{Mandel} indeed shows an example of non-lacunar
fractal, the Besicovitch fractal, which in modern terminology belongs to the
class of multinomial multifractals.  Mandelbrot \cite{Mandel} is actually
concerned about the low perceived lacunarity of the galaxy distribution and
how to reconcile it with the power-law galaxy correlation function.  In our
present study of cosmic voids, a non-lacunar fractal indeed appears as the
most suitable model.

At any rate, most fractal models of the galaxy distribution proposed so far
are lacunar and imply a self-similar distribution of cosmic voids.  The
self-similarity of voids has been considered by Einasto et al \cite{Einasto}
as a probe for scale invariance in the large scale structure.  Following the
ideas of Mandelbrot \cite{Mandel}, we have established that the rank-ordering
of fractal void sizes fulfills Zipf's power law and tried to confirm it with
data from galaxy surveys \cite{Gaite}.  Thus far, the scaling of galaxy voids
remains moot: our analysis does not show any evidence \cite{Gaite}, but
analyses of recent surveys are more favourable \cite{Tikho-Kara,Tikho,Tikho2}.
However, it is questionable that these scalings hold in a sufficiently long
range.

In a multifractal geometry, it is natural to define voids as the locations of
mass depletions \cite{I4}. This notion of voids is more general than the
notion of voids as {\em empty} holes and, actually, allows us to define voids
in non-lacunar fractals.  We study here the geometry of multifractal voids,
which is connected with the geometry of under-densities in a Gaussian field
but is actually more complicated.  From the multifractal geometry of dark
matter voids, we can derive the geometry of galaxy voids with a model of
galaxy biasing.  We consider a simple biasing model, inspired by the ``peak
theory'' of Gaussian fields \cite{Kaiser}, but we substitute Gaussian peaks by
multifractal mass concentrations, in accord with the multifractal model of
dark matter halos that we have proposed in Refs.~\cite{I,I4}.  Our model
provides a new perspective on Peebles' ``void phenomenon'' \cite{Pee2},
regarding the emptiness of voids.

We begin with the probabilistic analysis of voids in point distributions. We
calculate the probability of spherical voids and the size of the largest void
in a Poisson distribution (Sect.\ \ref{Poisson}).  We proceed to the general
probability of voids in correlated point distributions (Sect.\ \ref{correl})
but we obtain less complete results.  We introduce the hierarchical Poisson
model and the lognormal model as typical distributions and we connect them
with fractal distributions.  To measure the statistics of spherical voids in
samples, we devise an efficient void-finder in Sect.\ \ref{void-f}. In
Sect.~\ref{scaling}, we adopt a geometrical viewpoint centered on fractal
distributions and, in particular, cut-out sets, which we consider as cosmic
web models.  We generalize this construction to multifractals and thence we
proceed to a general study of multifractal voids, which leads us to
differentiate two types of voids according to their geometry (Sect.\
\ref{MF}). 
This general study of multifractal voids is combined in Sect.\ref{sim-gal}
with the results of Sect.~\ref{correl} to characterize the voids in
cosmological simulations and galaxy samples In particular, we deduce some
properties of galaxy voids in Sect.\ \ref{bias}, assuming a multifractal model
of galaxy biasing.  We summarize our results and present our conclusions in
Sect.\ \ref{discuss}.  Finally, we include three appendices to deal with
technical points.

\section{Poissonian analysis of cosmic voids}
\label{Poisson}

\subsection{The probability of voids}
\label{Poisson-p}

Let us recall the formulation of cosmic voids by Otto et al.\ \cite{Otto},
based on the results of Politzer \& Preskill \cite{Pol}.  They first quantify
the fluctuations in a sample of randomly distributed points (a homogeneous
Poisson field).  If the density of points is $n$, then the probability of
having $k$ points in a region of volume $V$ is given by the familiar Poisson
distribution with parameter $N = nV$:
$${P}_k[V] = \frac{N^k}{k!}\,e^{-N}.$$
The condition $k \ll N$ means that the given region is devoid of points.
However, the calculation of the probability of having a void of given size and
shape at any place is difficult.  Politzer \& Preskill's analysis \cite{Pol}
yields the following formula for the probability per unit total volume that
there is some region of volume $V$ and given shape that contains $k$ points:
$$
\mathbb{P}_k[V] = c_s\,
\frac{(nV)^3}{V}\,{P}_k[V]\,,
$$ 
where $c_s$ is a coefficient that depends on the shape.
To prove this formula, they simulate the action of a void-finding
algorithm. Thus, they calculate $c_s$ for spherical voids.

The simplest shape of a void is certainly the spherical shape.  Let us call
$\mathbb{P}_k[V]$ the probability density of having a volume $V$ spherical
void containing $k$ points.  A sphere containing $k$ points is defined by four
(non-coplanar) points on its boundary, because the sphere can always be
enlarged so as to touch four points. We call it a $k$-void, understanding that
$k \ll nV$.  $\mathbb{P}_k[V]$ can be estimated from a sample of the Poisson
distribution by counting the total number of $k$-voids and, then, dividing the
number of them that have volume between $V$ and $V + dV$ by their total
number.  To calculate the number of $k$-voids and $\mathbb{P}_k[V]$
analytically, we can employ the following method.

Since the points are uncorrelated, the probability distribution for each point
is constant and independent of the other points.  The number density of point
quadruplets is simply the product
$$
\frac{n^4}{4!}\,d^3x_1 \,d^3x_2\,d^3x_3\,d^3x_4\,,
$$ 
where the denominator takes into account that the points are unordered.
Its integral over the total volume is the total number of quadruplets $N_{\rm
t}(N_{\rm t}-1)(N_{\rm t}-2)(N_{\rm t}-3)/4! \approx N_{\rm t}^4/4!$, being
$N_{\rm t}$ the total number of points.
We can calculate the number density of $k$-voids as the product of the total
density of quadruplets and the probability ${P}_k[V]$, where $V$ is the volume
of the sphere that corresponds to a given quadruplet.
Therefore, the only problem is to express the four-point volume element in
terms of a set of variables that includes the volume of that sphere.  If we
denote the position of the center of the sphere (the quadruplet's
circumcenter) by $x_{\rm c}$, then the required number density is
\begin{eqnarray}
\frac{n^4}{4!}\,{P}_k[V]\,
d^3x_1 \,d^3x_2\,d^3x_3\,d^3x_4 = 
\nonumber\\ 
\frac{n^4}{4!}\,d^3x_{\rm c}\,{P}_k[V]\,V^2\,dV\,
f({\bm \th}_1,{\bm \th}_2,{\bm \th}_3,{\bm \th}_4)\,
d^2{\bm \th}_1\,d^2{\bm \th}_2\,d^2{\bm \th}_3\,d^2{\bm \th}_4,
\label{vol-elem}
\end{eqnarray}
where $\{{\bm \th}_i\}_{i=1}^4$ are the four sets of two angular coordinates
over the sphere, and $f$ is a function of these angular coordinates.  This
expression follows from translation invariance and dilation covariance only.
The function $f$ is calculated explicitly in Appendix \ref{nb-voids}.

To calculate the number of $k$-voids of volume $V$, we integrate the
right-hand side of Eq.~(\ref{vol-elem}) over all variables except $V$. If we
further integrate over $V$, we obtain the total number of $k$-voids. The
integral of $n\,d^3x_{\rm c}$ can be factored out, making the number of
$k$-voids proportional to the total number of points $N_t$.  The integral of
$f$ over the angular coordinates yields the factor $4!\,12\pi^2/35$ (see
Appendix \ref{nb-voids}). Therefore, the number of $k$-voids with volume
between $V$ and $V+dV$, per point of the sample, is
$$
\frac{12\,\pi^2}{35}\,{P}_k[V]\,n^3\,V^2\,dV\,,
$$
and the total number of $k$-voids per point is
$$
\frac{12\,\pi^2}{35}\int_{0}^{\infty}\frac{N^k}{k!}\,e^{-N} N^2\,dN =
\frac{12\,\pi^2}{35}\,(k+2)(k+1)\,.
$$ 
In particular, the number of voids (0-voids) per point is $24\pi^2/35 =
6.76773$.  It is interesting to compare this number with the numbers that
correspond to regular lattices.  For example, we can consider a body centered
cubic lattice, which has six void quadruplets (tetrahedra) per point
\cite{Auren}.  There are regular lattices with more voids per point, so it is
not surprising that the value in the random case is larger than six.

To obtain the probability of $k$-voids $\mathbb{P}_k[V]$, we divide the number
of $k$-voids of volume $V$ by the total number of $k$-voids:
\begin{equation}
\mathbb{P}_k[V] = \frac{1}{(k+2)(k+1)}\,
\frac{(nV)^3}{V}\,{P}_k[V]\,.
\label{cPP}
\end{equation}
Note that the name $k$-void makes sense only if $V \gg k/n$, but 0-voids are
voids for any $V$.  ${P}_0[V]$ approaches unity in the limit $V \ra 0$, but
$\mathbb{P}_0[V]$ is small in this limit, due to the boundary factor
(corresponding to the four points defining the void). In other words, the
probability that a randomly chosen small ball be empty is large, but the
probability that it have four points on its boundary and thus constitute a
void is small. In fact, the most probable void size (the mode) is $V=2/n$.  We
can calculate other statistical quantities, for example, the mean void size
${\overline V}=3/n$.

We can generalize the above method to other void shapes, e.g., ellipsoidal
voids, and indeed to any shape defined by a finite number of points.  The
number of $k$-voids has an expression generalizing Eq.~(\ref{vol-elem}), where
the exponent of $V$ is the number of points minus two; e.g, an ellipsoid is
defined by nine boundary points.  The number of $k$-voids is always
proportional to the number of sample points $N_{\rm t}$.  However, the
function $f$ of angular coordinates can be very complicated, and we may not be
able to integrate it and, thus, obtain the proportionality constant.
Nonetheless, $\mathbb{P}_k[V]$ is a power of $V$ with the given exponent, and
the normalization constant is easy to obtain.

Otto et al.\ \cite{Otto} applied their results to determining the significance
of some large voids that had been found at the time and concluded that those
voids were not significant, namely, that they did not rule out a Poisson
distribution. The conclusion that large voids are not significant was
contradicted by Betancort-Rijo \cite{Bet} and does not hold anymore, regarding
recent galaxy surveys. One recent survey is studied in Sect.~\ref{bias}.  In
the next section, we study in detail the statistics of large voids in Poisson
distributions.

\subsection{Extreme values: the largest and smallest voids}
\label{xtreme-P}

Following Otto et al.  \cite{Otto}, we consider the size of the largest void
in a given sample as the most important statistic. Indeed, the largest voids
are the ones to be first perceived.
In general, the statistics of extreme values is useful for data that are
naturally rank ordered. The largest value is the most important one.  The
smallest value and, hence, the total range can also be useful statistics, for
example, in a list of voids obtained with a void-finder.  The theory of
extreme values is a classic subject in statistics. A brief introduction to it
is given by Sornette \cite{Sornette}.

The rank order is related to the cumulative distribution $\mathbb{P}_>[V] =
\int_V^\infty\mathbb{P}_0[v]\,dv$, which we often employ in this work. In an
$M$-sample of this distribution, namely, a sample of $M$ void sizes, the rank
of one of them is the number of values larger than or equal to it, that is,
the order in the sorted list $V_1 \geq \cdots \geq V_M$.  If we have several
$M$-samples available, we can obtain the distribution of sizes for each rank
and its average.  If $M=1$, we can adopt as average value the median $V_{\rm
med}$, such that $\mathbb{P}_>[V_{\rm med}] = 1/2$. In the general case of
$M$-samples, suitable average values are given by $\mathbb{P}_>[V_m] =
m/(M+1)$, where $m$ denotes the rank.  Thus, the largest value $V_1$ fulfills
$\mathbb{P}_>[V_1] = 1/(M+1)$ and the smallest one fulfills $\mathbb{P}_>[V_M]
= M/(M+1)$.

For spherical voids in the Poisson distribution, we can calculate
$\mathbb{P}_>[V]$ from Eq.~(\ref{cPP}) (with $k=0$):
\begin{equation}
\mathbb{P}_>[V] = \left(1+nV+\frac{(nV)^2}{2} \right) e^{-nV}\,.
\label{Pcum}
\end{equation}
The probable sizes of the largest and smallest voids can be obtained by
solving the corresponding equations. Although these equations involve the
total number of voids $M$ rather than the total number of points $N_{\rm t}$,
we know that $M=6.77\,N_{\rm t}$. Of course, we assume that this number is
large. Thus, the equation for the largest void is
\begin{equation}
\mathbb{P}_>[V_1] \approx \frac{(nV_1)^2}{2}\, e^{-nV_1} = \frac{1}{M}.
\end{equation}
Letting $N_1=nV_1$ denote the number of points that correspond to the
largest void volume, we are to solve the equation
$$
N_1 = \ln \frac{M}{2} + 2 \ln N_1\,.
$$
It can be solved iteratively, starting with $N_1=1$. We obtain the asymptotic
expansion 
\begin{equation}
N_1= \ln \frac{M}{2} + 2 \ln \ln \frac{M}{2} + \cdots
\label{N1}
\end{equation}
For a sample with $N_{\rm t}= 10\hspace{1pt}000$ points, 
$N_1 \simeq \ln 33\hspace{1pt}839 + 2 \ln \ln 33\hspace{1pt}839 \simeq 15.$

It is easy to generalize these results to non-spherical voids: in
Eq.~(\ref{Pcum}), the corresponding terms of higher degree are to be added to
the polynomial in its right-hand side. Consequently, the coefficient of the
sub-leading term in Eq.~(\ref{N1}) is larger. Besides, the coefficient of
proportionality between $M$ and $N_{\rm t}$ grows as well, but this has a
smaller effect.  Although the shape of voids is, in principle, a sub-leading
effect, the polynomial in the right-hand side of Eq.~(\ref{Pcum}) tends to
$e^{nV}$ as the number of points that define the shape of voids goes to
infinity and, therefore, $\mathbb{P}_>[V]$ tends to one. In other words, the
volume of the voids grows with no bound when we remove any constraint on their
shape, as is natural.  In Sect.~\ref{void-f}, we mention constraints that are
suitable for void-finding algorithms.

The equation for the smallest void is
\begin{equation}
\mathbb{P}_>[V_M] \approx \left(1+nV_M \right) e^{-nV_M} = \frac{M}{M+1}
\approx 1 - \frac{1}{M}\,,
\end{equation}
with solution
\begin{equation}
N_M = \frac{1}{\sqrt{M}}\,.
\label{N_M}
\end{equation}
For spherical voids in $N_{\rm t}= 10\hspace{1pt}000$ points, there are about
$67\hspace{1pt}677$ voids and the smallest void is such that
$N_{67\hspace{1pt}677} = 0.0038$.

We can estimate the probable volume of the largest void in another way, by
using $P_0[V]$, which is the probability that a {\em given} region of volume
$V$ be empty.  Given a sample with total volume $V_{\rm t}$, let us tile it
with $V_{\rm t}/V$ regions of volume $V$, for example, applying a cubic mesh
to it.  The expected number of empty regions is $P_0[V] (V_{\rm t}/V)$.
Therefore, if only one of them is to be empty, then we have the equation
$P_0[V] (V_{\rm t}/V)=1$ or
$$
e^{-N_1} = \frac{N_1}{N_{\rm t}}.
$$
This equation can also be solved iteratively, yielding
$$ N_1= \ln N_{\rm t} - \ln \ln N_{\rm t} + \cdots
$$
For $N_{\rm t}= 10\hspace{1pt}000$ points, now $N_{1} \simeq 7$.
Since $P_0[V]$ is the probability that a {\em given} region of volume $V$ be
empty, it represents a random trial region that is unlikely to be empty even
if it intersects the largest void of the same volume (as pointed out by Otto
et al.\ \cite{Otto}). Therefore, this method underestimates the size of the
largest void.  Nevertheless, the leading logarithmic term is the same as the
leading term in Eq.~(\ref{N1}), obtained from $\mathbb{P}_>[V]$.

By just taking the leading term $\ln N_{\rm t}$, we can deduce that the
relative size of the largest void, namely, the fraction of the sample's volume
that it occupies, vanishes as $N_{\rm t} \ra \infty$. In other words, voids
disappear in the continuum limit $n \ra \infty$ of a Poisson distribution, as
is natural.

\section{The probability of voids in correlated point distributions}
\label{correl}

It is possible to extend the preceding methods to correlated distributions of
particles. Otto et al.\ \cite{Otto} already derived one formula for the
probability of voids in correlated point distributions, in terms of an
expansion in powers of the density $n$.  Betancort-Rijo \cite{Bet} employed
instead a non-perturbative method based on the expression of $P_0[V]$ in terms
of the probability of density fluctuations.  We take inspiration from both
approaches to generalize the preceding calculations.  Ultimately, we intend to
find the behavior of the probability of voids in distributions with strong
correlations, namely, in distributions with strong mass concentrations.  In
particular, we study fractal distributions.

\subsection{The general probability of voids}
\label{gPvoids}

The probability of spherical voids in the general case is studied by Otto et
al.\ using expansions in powers of the density $n$ \cite{Otto}.  The function
${P}_0[V]$ (void probability function) was previously expressed by White
\cite{White} as
\begin{equation}
{P}_0[V] = \exp\left[\sum_{k=1}^{\infty} \frac{(-nV)^k}{k!}\, 
{\bar\xi}_k \right],
\label{vpf}
\end{equation}
where ${\bar\xi}_1 = 1$, and, for $k \geq 2$, ${\bar\xi}_k$ 
is the average in $V$ of the correlation function ${\xi}_k$:
$${\bar\xi}_k = \frac{1}{V^k} \int_V d^3x_1 \cdots d^3x_k\, \xi_k(x_1, \ldots,
x_k).$$ 
${P}_0[V]$ is analogous to the grand-canonical partition function in
statistical mechanics (with velocities integrated over) and the expansion in
terms of the ${\bar\xi}_k$ (the cumulants) is analogous to the cluster
expansion. This connection was noticed by Otto et al.\ \cite{Otto} and is
considered further by Mekjian \cite{Mekjian}.  We assume that the cumulants
${\bar\xi}_k$ ($k \geq 2$) always vanish in the limit $V \ra \infty$ and that
they vanish the more rapidly the larger $k$ is.  Thus, the first correction to
the Poisson formula is given by a non-vanishing ${\bar\xi}_2$ (a Gaussian
field).  Of course, Eq.~(\ref{vpf}) is only valid insofar as the expansion
converges
(see Sect.~\ref{scope}).

White \cite{White} noticed that ${P}_0[V]$ can also be expressed, using the
local particle density $\r$, as the integral over $\r$ of the product of the
probability of $\r$ and the probability of a void volume $V$ given that the
density is $\r$.  The latter probability is given by the Poisson
formula. Therefore,
\begin{equation}
{P}_0[V] = \int\limits_0^\infty P(\r)\,\exp\left[-\r V\right] d\r\,.
\label{vpf-int}
\end{equation}
The exponential inside the integral can be expanded in powers of $\r V$ and,
assuming that the integration and the sum can be interchanged, ${P}_0[V]$
becomes an expansion in terms of moments:
\begin{equation}
{P}_0[V] = \sum_{k=0}^{\infty} \frac{(-nV)^k}{k!}\, {\bar\mu}_k\,,
\label{vpfm}
\end{equation}
where the moments are defined by
$$
{\bar\mu}_k = \int\limits_0^\infty \left(\frac{\r}{n}\right)^k P(\r)\,d\r =
\frac{1}{(nV)^k} \int_V d^3x_1 \cdots d^3x_k\, \langle\r(x_1) \cdots
\r(x_k)\rangle,
$$ 
with $n=\langle \r \rangle$ (note that ${\bar\mu}_1 \equiv 1$).  Of course,
the moment expansion in Eq.~(\ref{vpfm}) is equivalent to the cumulant
expansion in Eq.~(\ref{vpf}).
However, the integral expression (\ref{vpf-int}) is more general
and is always valid (see Sect.~\ref{scope}) .  

If we remove the density fluctuations by making $P(\r) = \d(\r - n)$, we
recover the Poisson formula, that is, the first term in the
expansion~(\ref{vpf}). If we account for small fluctuations by taking a
Gaussian $P(\r)$ \cite{Bet}, we further obtain the second term in the
expansion~(\ref{vpf}).  We must also consider a variant of the Poisson
distribution, such that the total space is divided in two regions, one being
empty and another having the Poisson distribution. This second region is meant
to be formed by matter clusters (of uniform density).  If $n_\mathrm{c}$
denotes the density inside the clusters, we have that
$$P(\r) = \left(1-\frac{n}{n_\mathrm{c}}\right) \d(\r) + 
\frac{n}{n_\mathrm{c}}\, \d(\r - n_\mathrm{c}).$$
The two constants before the delta functions are deduced from the
normalization of $P(\r)$ and the condition $\langle \r \rangle=n$.
The void probability function given by Eq.~(\ref{vpf-int}) is
\begin{equation}
{P}_0[V] = 1- \frac{n}{n_\mathrm{c}}\left(1-e^{-n_\mathrm{c}V}\right) =
1 + \frac{n}{n_\mathrm{c}}\sum_{k=1}^{\infty} \frac{(-n_\mathrm{c}V)^k}{k!}\,.
\label{hP}
\end{equation}
Comparing it with Eq.~(\ref{vpfm}), we deduce that ${{\bar\mu}_2} =
n_\mathrm{c}/n$ and that the moments fulfill the hierarchical relation
${\bar\mu}_k = {{\bar\mu}_2}^{k-1}$.  This hierarchical model is especially
interesting when $n_\mathrm{c} \gg n$, namely, when the volume occupied by the
matter clusters is a small fraction of the total volume. Then, it coincides
with Fry's hierarchical Poisson model \cite{Fry}, with
\begin{equation}
{P}_0[V] = \exp\left[-\frac{1-e^{-nV{\bar\xi}_2}}{{\bar\xi}_2}\right]
\label{FhP}
\end{equation}
(note that then ${\bar\xi}_2 = {\bar\mu}_2 -1 \gg 1$).  Hierarchical models
are related to fractal models \cite{Borga_h,Borga}, which we study in detail
in Sect.~\ref{nonlin-scaling}.

The moments always satisfy the inequality ${\bar\mu}_k \geq
{{\bar\mu}_2}^{k-1}$, among other inequalities \cite{inequal}.  As we have
seen, the lowest allowed value of ${\bar\mu}_k$ occurs in distributions that
are uniform inside their support. In spite of this uniformity, there are
strong density fluctuations in the total volume when $n_\mathrm{c} \gg n$,
since the support is sparse. We now consider the lognormal model
\cite{Coles-Jones}, which can have strong density fluctuations inside its
support, such that ${\bar\mu}_k \gg {{\bar\mu}_2}^{k-1}$. This model is also
related to fractal models.

For the lognormal model,
\begin{eqnarray}
{P}_0[V] &=& \frac{1}{\s\sqrt{2\pi}}\int\limits_0^\infty 
\exp\left[-\frac{(\ln\r-\mu)^2}{2\s^2}-\r V\right]
\frac{d\r}{\r} \nonumber\\
&=& \frac{1}{\s\sqrt{2\pi}}\int\limits_0^\infty 
\exp\left[-\frac{(\ln r+\s^2/2)^2}{2\s^2}- r\, nV\right]
\frac{dr}{r}\,,
\label{vpf-logn}
\end{eqnarray}
where $\mu= \ln n -\s^2/2$ (with $n=\langle \r \rangle$) and we have made the
change of variable $r = \r/n$ to obtain the second integral.  We can restrict
the support of this distribution as well, with the consequent addition of a
constant to ${P}_0[V]$ and substitution of $n$ by $n_\mathrm{c}$. But we
understand that the support is the total volume, for the moment.  The constant
$\s$ in Eq.~(\ref{vpf-logn}) actually depends on $V$, and $P(\r)$ becomes
Gaussian and eventually uniform as $V$ grows and $\s \ra 0$.
However, if we take $\s \simeq 1$, then the probability that a relatively
large volume be empty, namely, that $V \gg n^{-1}$, is much larger that the
corresponding Poisson value.  For example, with $\s = 1$ and $nV = 10$,
Eq.~(\ref{vpf-logn}) yields ${P}_0[V] = 0.0539$, whereas $\exp(-10) =
0.0000454$.

The lognormal moments are easily evaluated, yielding ${\bar\mu}_k =
\exp\left[k(k-1)\s^2/2\right]$.  In the limit $\s \ra 0$, ${\bar\mu}_k$ tends
to one and ${\bar\xi}_k$ vanishes (if $k \geq 2$). Moreover, the cumulants
actually vanish the more rapidly the larger $k$ is.  In the opposite limit,
when $\s$ is large, the moments grow exponentially and, furthermore,
${\bar\mu}_k \gg {{\bar\mu}_2}^{k-1}$.

\subsubsection{Calculation of the distribution of voids}

To calculate the number of spherical voids in a correlated point distribution,
the expression for a quadruplet of uncorrelated points in the left-hand side
of Eq.~(\ref{vol-elem}) must be replaced by
\begin{eqnarray}
\frac{n^4}{4!}\,
d^3x_1 \cdots d^3x_4 
\left[ {\z_1}(x_1) {\z_1}(x_2) {\z_1}(x_3) {\z_1}(x_4) + 
\right.\nonumber\\ \left.
{\z_2}(x_1,x_2) {\z_1}(x_3) {\z_1}(x_4) + \cdots +
{\z_1}(x_1) {\z_1}(x_2) {\z_2}(x_3,x_4) + 
\right.\nonumber\\ 
{\z_2}(x_1,x_2) {\z_2}(x_3,x_4) +
\cdots +
{\z_2}(x_1,x_4) {\z_2}(x_2,x_3) +
\nonumber\\
{\z_3}(x_1, x_2 ,x_3) {\z_1}(x_4) +
\cdots + {\z_1}(x_1) {\z_3}(x_2,x_3,x_4) + 
\nonumber\\
\left. 
{\z_4}(x_1, \ldots ,x_4) \right] {P}_0[V]\,,
\label{Pclus}
\end{eqnarray}
where the ${\z}(\cdot)$ are correlation functions conditioned by the presence
of the void (using White's general definitions \cite{White}).  These
correlation functions can be expanded in powers of $n$ and written in terms of
ordinary correlation functions:
\begin{eqnarray*}
{\z_1}(x) &=& 1 + 
\sum_{k=1}^{\infty} \frac{(-n)^k}{k!}
\int_V d^3y_1 \cdots d^3y_k\, 
\xi_{k+1}(x,y_1, \ldots, y_k),\\
{\z_2}(x_1,x_2) &=& {\xi}_2(x_1,x_2) +\\
&&\sum_{k=1}^{\infty} \frac{(-n)^k}{k!}
\int_V d^3y_1 \cdots d^3y_k\, 
\xi_{k+2}(x_1,x_2,y_1, \ldots, y_k),
\end{eqnarray*}
etc.  In particular, ${\z_1}(x)$ is independent of $x$, due to translation
invariance, and $n\,{\z_1}$ can be interpreted as a conditional density.
From expression (\ref{Pclus}), we can proceed like in the Poisson case, namely,
we can express the four-point volume element in terms of the sphere's volume
and angular variables,
factor out $n\,d^3x_{\rm c}$, and then integrate over the angular coordinates.
Unfortunately, this integration cannot be explicitly worked out in the general
case, because the ${\z}(\cdot)$ depend on the angular coordinates.

Naturally, an expression for the number of non-spherical voids, e.g.,
ellipsoidal voids, can also be written; but it is even less tractable than the
expression for spherical voids.  On the other hand, we could consider
$k$-voids instead of empty voids. Needless to say, the extra $k$ points also
make the expressions less tractable.

In one dimension, the geometry is much simpler. In general, only the two-point
function is involved in the expression corresponding to (\ref{Pclus}), there
are no angular coordinates, and there is one void interval per point.
Therefore,
\begin{equation}
\mathbb{P}_0[L] = 
\frac{n}{2}\,[\z_1^2+\z_2(L)]\,{P}_0[L].
\label{pvoids-1d}
\end{equation}
Of course, this one-dimensional formula is not directly applicable to
three-dimensional cosmic voids, but it is applicable to void intervals in
pencil-beam surveys, as measured by Einasto et al \cite{Einasto}, for example.

\subsection{Scope of the perturbative expansions in the density}
\label{scope}

The expansions in powers of the density used in the preceding section
require that the density be small, in a sense that we need to make precise.
The simplest expansion to study is the cumulant expansion of $P_0[V]$ given by
Eq.~(\ref{vpf}).  
The successive terms of this series are expected to be of small magnitude and
eventually decreasing.  In particular, the second term of the series is
smaller than the first (Poisson) term if $N = nV < {\bar\xi}_2^{-1}$. We can
interpret this condition in terms of the number variance in the volume $V$,
namely,
$$
\frac{\langle \d\! N^2 \rangle}{N^2} = \frac{1}{N} + {\bar\xi}_2\,.
$$
When $N < {\bar\xi}_2^{-1}$ the Poisson fluctuations dominate over the
fluctuations due to correlations, and vice versa.
However, the cumulant expansion~(\ref{vpf}) can converge in spite of having
initial terms that increase. For example, the expansion of $P_0$ for Fry's
hierarchical Poisson model in Eq.~(\ref{FhP}) converges for any value of $N
{\bar\xi}_2$, although the initial terms increase in magnitude when $N
{\bar\xi}_2 > 1$, until the $k$th term is such that $k \approx N
{\bar\xi}_2$. Thus, when $N {\bar\xi}_2 \gg 1$, it is necessary to sum a large
number of terms, some of which are large (in absolute value), to obtain a
negligible sum, for $e^{-N{\bar\xi}_2} \ra 0$ as ${N{\bar\xi}_2} \ra
\infty$. This makes the series unsuitable for numerical computations.

The cumulant expansion of the lognormal model has worse behavior.  Indeed, the
integral formula (\ref{vpf-logn}) shows that $P_0[V]$ is not analytic at $V=0$
\cite{Coles-Jones}. In consequence, the moment expansion~(\ref{vpfm}) must
diverge.  This also follows from the expression ${\bar\mu}_k =
\exp\left[k(k-1)\s^2/2\right]$.  $\ln P_0[V]$ is also not analytic at $V=0$
and the cumulant expansion diverges as well.  Actually, the moment and
cumulant expansions are asymptotic as $N \ra 0$.  They also are alternating
series. An alternating asymptotic series converges while its terms are of
decreasing magnitude, and it must be terminated at the term with the least
magnitude (which should be included with half its value).  Therefore, a small
value of $N{\bar\xi}_2$ is mandatory.  We can relate the bad behavior of these
density expansions in the case of strong clustering ($\s \gg 1$) to the strong
growth of moments, such that ${\bar\mu}_k \gg {{\bar\mu}_2}^{k-1}$. This
strong growth of moments is typical of certain distributions, in particular,
multifractals.

For the lognormal void probability function at large $\s$, the asymptotic
expansion of the integral in Eq.\ (\ref{vpf-logn}) worked out in Appendix
\ref{asymptotia} yields
\begin{equation}
P_0[V] = 1 - \frac{\sqrt{2 N}}{\s}\, e^{-\s^2/8} \left[1 + 
{\Mfunction{O}(\s^{-2})}\right].
\label{asy-vpf}
\end{equation}
This expression provides a reasonable approximation already for moderately
large $\s$. For $\s=3$, the approximation is quite good in a relevant range of
$N$. When $N=1$, the value computed directly with Eq.\ (\ref{vpf-logn}) is
$P_0[V] = 0.885346$ and the one computed with Eq.\ (\ref{asy-vpf}) is $P_0[V]
= 0.846957$. In contrast, the condition $N < {\bar\xi}_2^{-1} \simeq
\exp(-\s^2) = 0.00012$ is far too restrictive.

The non-analyticity of the lognormal $P_0[V]$ at $V=0$ is due to the slow
decay of its probability function $P(\r)$ in the high density limit.  In
general, mass distributions that are singular on small scales possess
probability functions with fat tails in the high density limit.  Nevertheless,
the probability function $P(\r)$ can have moments of any order, like the
lognormal distribution. Then, the series expansions of $P_0[V]$ or
${\z}(\cdot)$ are well defined, but they are asymptotic rather than
convergent.  Distributions with moments of any order but with fat tails such
that moment and cumulant expansions diverge can be called lognormal like.  In
these distributions, the full set of moments or cumulants does not determine a
unique function $P(\r)$.  To do this, one needs to add moments of non-integer
order and even moments of negative order.  In particular, the description of
multifractals requires all the $q$-moments, with $q$ any real number
\cite{Borga,Falc}

\subsection{The largest void in a correlated distribution}
\label{xtreme-c}

In absence of the probability function $\mathbb{P}_0[V]$, we do not have
detailed knowledge of the distribution of void sizes, but we can always find a
lower estimate of the largest void size by using ${P}_0[V]$ as in
Sect.~\ref{xtreme-P}: a lower estimate of the largest void size is given by
the solution of the equation $P_0[V] (V_{\rm t}/V)=1$.  For example, we
consider the hierarchical Poisson model, with $P_0[V]$ given by
Eq.~(\ref{hP}), and the lognormal $P_0[V]$ in Eq.\ (\ref{vpf-logn}).  For the
largest void, $N = nV \gg 1$, in general. Thus, we must focus on this limit,
which we can interpret, alternately, as the limit of large volumes or as the
continuum limit.

In the hierarchical Poisson model, we can take the continuum limit $n
\ra\infty$ while preserving the distribution of clusters, just by keeping
${{\bar\mu}_2} = n_\mathrm{c}/n$ constant. Thus, the voids inside clusters
vanish while the voids between clusters are unaffected. In that limit,
Eq.~(\ref{hP}) becomes
\begin{equation}
{P}_0[V] = 
1- \frac{n}{n_\mathrm{c}} = 1- \frac{1}{{\bar\mu}_2}\,.
\label{hP-lim}
\end{equation}
This quantity is close to one when the clustering is strong, namely, when
${{\bar\mu}_2} \gg 1$.  Then, the fraction of empty space is large, even in
the continuum limit of the distribution, and naturally we can find large
voids.  The largest void size is determined by the distribution of clusters,
which is assumed to be a Poisson distribution.

Regarding the lognormal model, we need the asymptotic form of the integral in
Eq.\ (\ref{vpf-logn}) for large $N=nV$ that is worked out in Appendix
\ref{asymptotia}.  Keeping only the first term, we have
\begin{equation}
\ln P_0 \approx -\frac{(\ln N)^2}{2\,\s^2}\,.
\label{asy_P0}
\end{equation}
In this case, $\lim_{N \ra\infty} P_0 = 0$, like in the Poisson distribution.
Naturally, the lognormal $P_0[V]$ has a decrease with $V$ that is slower than
the exponential decrease of the Poisson distribution.  For the largest void,
we have the equation:
$$
-\frac{(\ln N)^2}{2\,\s^2} = \ln\frac{N}{N_{\rm t}}\,.
$$
Its solution is simply
\begin{equation}
\ln N \approx \s \sqrt{2\ln N_t}\,.
\label{N1-c}
\end{equation}
Compared to the leading asymptotic term in the Poisson case, namely, $N = \ln
N_{\rm t}$, we notice that the size of the largest void is greatly enhanced.
For example, taking $N_{\rm t}= 10\hspace{1pt}000$ points and $\s = 1$, now $N
\simeq 73$.  Nevertheless, the relative size $V/V_{\rm t}$ of the largest void
goes to zero as $N_{\rm t} = nV_{\rm t} \ra \infty$, like in the Poisson case.

A remark is in order here: $\s$ is not really a constant, but decreases with
$V$, as we comment after Eq.~(\ref{vpf-logn}). This fact can be taken into
account to derive an equation for the size of the largest voids more accurate
than Eq.~(\ref{N1-c}) if we know the law $\s(V)$. The scaling model that we
study in Sect.~\ref{nonlin-scaling} provides us with a simple form of that
law. At any rate, Eq.~(\ref{N1-c}) is sound if $\s$ is of the order of unity
at the scale $V$ of the largest void.

In general, the largest value in a random sample from a distribution is
determined by the tail of the distribution. If the sample is very large, the
distribution of the largest value must approach one of three distributions,
named extreme value distributions \cite{Sornette}. Only two extreme value
distributions are relevant in our context: the Gumbel distribution and the
Fr\'echet-Pareto distribution (the third distribution is for bounded
variables).  The domain of attraction of the Gumbel distribution consists of
the probability density functions with a tail falling faster than a power law.
It contains the Poisson distribution of voids and also lognormal-like void
probability functions.  The domain of attraction of the Fr\'echet-Pareto
distribution consists of the probability density functions with a power law
tail. The Fr\'echet-Pareto distribution assigns considerable probability to
large values and is such that the largest value is of the same order of
magnitude as the sum of the remaining values, even as the number of values
tends to infinity.  We have found an example of this behavior in Fry's
hierarchical Poisson model.  Voids of this type are typical of fractal
distributions, which we study next.

\subsection{Voids in scaling distributions}
\label{nonlin-scaling}

We have seen that the analytical description of voids is difficult when the
density of points is large or the correlations are strong, namely, when $N
{\bar\xi}_2 > 1$.  Of course, the analytical treatment of strongly correlated
distributions, with ${\bar\xi}_2 \gg 1$, is difficult in general.  Some
progress can be made with the use of tractable specific models which,
arguably, have general features. For example, we have mentioned that the
Gumbel or Fr\'echet-Pareto distributions are limit distributions of the
largest void, featuring a limited or unlimited size of it, respectively.  As
specific models, we have chosen the hierarchical Poisson model and the
lognormal model, connected with the Fr\'echet-Pareto and Gumbel distributions,
respectively. Our two specific models are also connected with fractal
distributions, which have special interest.

We can introduce self-similar fractals as a generalization of the hierarchical
Poisson model. This hierarchical model has only two levels in the hierarchy,
namely, clusters and particles; but we can imagine that every particle is also
a cluster at the adequate level of resolution, thus building a hierarchy with
many levels, even an infinite number.  An infinite self-similar cluster
hierarchy constitutes a fractal.  More formally, a fractal is a continuous
distribution (with mean density $n \ra \infty$) such that it has strong
correlations with scaling properties; namely, the correlation functions
${\xi}_k$ as well as their integrals ${\bar\xi}_k$ are power laws.  In
particular, the two-point correlation is ${\xi}_2(r) = (r_0/r)^{\g}$, where
$r_0$ is the homogeneity scale, and the relative mass variance in a cell of
volume $V$ is ${\bar\xi}_2 = (V_0/V)^{\g/3}$, where $V_0$ is a 
homogeneity volume (of the order of magnitude of $r_0^3$).

We must distinguish monofractals, whose scaling properties are characterized
by just the exponent $\g$ or the dimension $D = 3-\g$, from multifractals,
which have a spectrum of dimensions.  Fractal distributions can have voids in
spite of being continuous and statistically homogeneous,%
\footnote{Statistical homogeneity means that statistical quantities are
  translation invariant and, of course, is weaker than strict homogeneity.}
 unlike other types of distributions. For example, the Poisson distribution 
or lognormal-like distributions do not have voids in the continuum limit.
On the other hand, when the density $n$ is finite, the
fractal regime holds while $nV{\bar\xi}_2>1$, namely, for volumes $V$ such
that $n\,V_0^{\g/3}\, V^{D/3}>1$.

Since a fractal distribution is scale invariant, it is reasonable to assume
that the distribution of voids in a fractal is also scale invariant. Before
considering the scaling of voids in three dimensions, let us consider
one-dimensional fractals, in which the geometrical problems regarding 
the shape of voids are absent.

In one dimension, a probabilistic formulation of the scaling of voids is
provided by the application of the L\'evy {\em stable} distributions
\cite{Mandel}.  Their stability means that the sum of a number of independent
identical variables keeps the same distribution, after rescaling.  A L\'evy
distribution becomes a power law for sufficiently large values of its variable
and, furthermore, it is an attractor of distributions with the same power-law
behaviour (generalizing the central limit theorem).  Therefore, a distribution
such that the interval between successive points has probability
$\mathbb{P}_0[L] \sim L^{-D-1},$ for $0<D<1$ and large $L$, converges in the
continuum limit to a {\em L\'evy flight} (a generalized random walk), in which
$\mathbb{P}_0[L] \propto L^{-D-1},$ for any $L$, and the mass in an interval
of length $l$ is distributed as $l^D$.  This power-law form of
$\mathbb{P}_0[L]$ is not integrable at $L=0$ and must be understood as a
conditional probability: the probability of a void of length $L$ given that
$L$ is longer than an arbitrary length $\l$ is the Pareto distribution
$\mathbb{P}_0[L | L > \l] = D\l^D/L^{D+1}$ (see Ref.~\cite{Mandel} and
Appendix \ref{L-flight}).

Thus, the construction of a L\'evy flight directly defines $\mathbb{P}_0[L]$,
without resort to expression (\ref{pvoids-1d}) for one-dimensional voids.  In
particular, the void probability function $P_0[L]$ tends to one and becomes
irrelevant for the probability of voids.  In fact, $P_0[L] \ra 1$ as a
consequence of the strong correlations between the points in a L\'evy flight:
a {\em given} interval is likely to contain no points, because the probability
of finding one point is only non-negligible when conditioned on being close to
another point.  Mandelbrot \cite{Mandel} shows that the conditional
probability of having some mass in an interval of length $l$ inside an
interval of length $L$ that is known to have mass is $P[M(l) >0 | M(L) >0] =
(l/L)^{1-D}$. Hence, $P_0[L]$ is not exactly one due to the existence of the
homogeneity scale $r_0$, and we can write
$$P_0[L] = 1 - P[M(L) >0 \,| M(r_0) >0] = 1 - (L/r_0)^{1-D},$$ 
which tends to one as $L/r_0 \ra 0$.  Remarkably, this holds {\em
independently} of the value of the density $n$, even as $n \ra \infty$.

Both properties, namely, $\mathbb{P}_0[L] \propto L^{-D-1}$ and $P_0[L] \ra 1$
for $L \ll r_0$, can be generalized to other one-dimensional fractals (with
voids that are not independent) and also to fractals in higher dimensions.
Actually, higher-dimensional L\'evy flights are not useful to model the
scaling of voids, because in them an empty interval between two points does
not define a void.  Therefore, it is necessary to employ more elaborate
geometrical concepts.  We introduce these concepts in Sect.~\ref{scaling} and,
thence, we formulate the appropriate version of the power-law form of the
probability of voids $\mathbb{P}_0[V]$.  In dimension $d$ and letting $V$
denote the void's $d$-volume, we can loosely write this law as
$\mathbb{P}_0[V] \sim V^{-D_{\rm b}/d-1}$, where $D_{\rm b}$ is the {\em
box-counting dimension} (which is equal to $D$ for L\'evy flights).  The
behavior of $P_0[V]$ for small $V$ in a fractal is also governed by $D_{\rm
b}$, because, by definition, the number of non-empty boxes of size $V$ follows
the power law $V^{-D_{\rm b}/d}$ \cite{Falc}.  Therefore, the ratio of
non-empty boxes, which we can interpret as the probability that one box be
non-empty, follows the power law $V^{1-D_{\rm b}/d}$.  In consequence,
\begin{equation}
P_0[V] = 1 - (V/V_0)^{1-D_{\rm b}/d}, 
\label{fractal-vpf}
\end{equation}
where $V_0$ is the homogeneity volume.  If $D_{\rm b} < d$, $P_0[V]$ tends to
one as $V \ra 0$. Conversely, as $V \ra V_0$, $P_0[V]$ vanishes. That is to
say, $V_0$ represents the size of the largest voids.

The condition $D_{\rm b} < d$ is fulfilled by L\'evy flights and, in general,
by monofractals, since they have only one dimension $D$, which has to be
smaller than $d$.  In this regard, note that Eq.~(\ref{hP-lim}) for the
hierarchical Poisson model coincides with the particular case of
Eq.~(\ref{fractal-vpf}) in which $D_{\rm b} = d - \g$ and ${\bar\mu}_2 =
(V_0/V)^{\g/d}$.  However, a {\em multifractal} can have box dimension $D_{\rm
b} = d$ while its other dimensions are smaller than $d$.  Then, $P_0[V]$ does
not tend to one for small $V$ and actually vanishes, according to
Eq.~(\ref{fractal-vpf}).  These multifractals belong to the type of {\em
non-lacunar} fractals introduced by Mandelbrot \cite{Mandel}.  We can obtain
more information about $P_0[V]$ by recalling that $P(\r)$, in this case, can
be expanded around the value that maximizes $\r$ (the mode), resulting in a
lognormal distribution \cite{JonesCM,I4}.  Its void probability function is
given by Eq.~(\ref{vpf-logn}), which has the asymptotic form given by
Eq.~(\ref{asy_P0}) and, therefore, vanishes in the continuum limit $n \ra
\infty$.

Besides, it is instructive to compare the asymptotic form of the lognormal
$P_0[V]$ in the nonlinear regime given by Eq.~(\ref{asy-vpf}) with the form
given by Eq.~(\ref{fractal-vpf}) for fractals with $D_{\rm b} < d$.  Both
equations imply that $P_0[V]$ approaches one if the correlations are strong,
namely, if $V \ll V_0$ in Eq.~(\ref{fractal-vpf}) or if $\s$ is large in
Eq.~(\ref{asy-vpf}). However, the condition is independent of the density in
Eq.~(\ref{fractal-vpf}), whereas $N = nV$ appears in Eq.~(\ref{asy-vpf}).
Thus, in this equation, $P_0[V]$ is close to one only as long as $N \ll \s^2
\exp(\s^2/4)$ (a value that can be large), but the equation becomes invalid
for larger $N$ and, in fact, $P_0[V] \ra 0$ as $N \ra \infty$.
Multifractals with $D_{\rm b} = d$ have {\em no voids} in the continuum limit,
like the Poisson distribution, but the large fluctuations due to the strong
correlations imply that volumes such that $N$ is relatively large still have a
good chance of being empty, unlike in the Poisson distribution.

To summarize, let us restrict ourselves to $d=3$: fractals with $D_{\rm b} <
3$ have scaling voids whereas multifractals with $D_{\rm b} = 3$ are
non-lacunar fractals and do not have voids at all in the continuum limit
(assuming that they are statistically homogeneous).  Finite samples of these
non-lacunar fractals have voids, but $\mathbb{P}_0[V]$ does not have to be a
power law.  To verify these conclusions, we resort to geometric methods in
Sects.~\ref{scaling} and \ref{MF}. In particular, we show in Sect.~\ref{MF}
that voids of a more sophisticated type can actually be defined in
multifractals with $D_{\rm b} = 3$.

\section{A spherical void finder}
\label{void-f}

We have seen that it is very difficult to derive a general analytic expression
of the probability of spherical voids $\mathbb{P}_0[V]$.  However, the Poisson
law given by Eq.~(\ref{cPP}) or the Pareto law in fractals with $D_{\rm b} <
3$ are very suitable for experimental confirmation.  The experimental
confirmation can be achieved through the measure of the statistics of voids in
simulated samples.  In fact, the statistic most easily obtained from
simulations is the rank order of void sizes.  The rank-ordering corresponding
to the Pareto law is known as Zipf's law.

The voids in a point distribution can be extracted with the help of a suitable
void-finder.  Many void-finders have been devised already, which essentially
differ in the definition of voids that they use. A comparison of void-finders
is made by Colberg et al \cite{Col_etal}.  The more recent void-finders
usually define voids of variable shape \cite{Kauffmann,ElAd-Pir,Tikho}.  We
have devised a finder of variable-shape voids based on the Delaunay
tessellation of a set of points \cite{I2}. This tessellation is the natural
geometric construction to use, for it provides the {\em unique} set of empty
spheres associated with the set of points, in the sense that each sphere is
defined by four non-coplanar points that form a Delaunay simplex \cite{Auren}.
Thus, these spheres are precisely the spherical voids defined in
Sect.~\ref{Poisson-p}, which the algorithm of Ref.~\cite{I2} merges if they
have sufficient overlap.  The overlap is measured by a suitable parameter.
The sequences of voids found in simulated monofractals follow Zipf's law
\cite{I2}. However, one must avoid too much merging of spheres, for the voids
can adopt too complex shapes and, sometimes, one void percolates through the
sample.

In general, finders of variable-shape voids must have adjustable parameters to
prevent inappropriate shapes such as the dumb-bell shape, as discussed by ElAd
and Piran \cite{ElAd-Pir}. We remark in Sect.~\ref{xtreme-P} that the absence
of any restriction on the shape of voids actually leads to the presence of
unbounded voids even in a Poisson distribution. This is intuitively obvious,
for the void can then snake through the set of points. This snaking can be
averted just by demanding that voids be convex. Unfortunately, this neat
condition is very complex from the algorithmic standpoint.

The simplest way to forbid strange shapes of voids is to prescribe voids of
constant regular shape.  We have already shown that void-finders based on
voids of constant shape are suitable for demonstrating the scaling of voids in
fractal distributions \cite{Gaite}. And the simplest shape is the sphere.  Of
course, the natural set of spherical voids is the set of spheres defined by
the Delaunay tessellation, which we do not need to merge.  It is natural to
require that the spheres are contained in the sample region and that they do
not overlap. This condition removes some spheres, but we always want to keep
the largest one.  Thus, a convenient algorithm begins by finding the largest
sphere contained in the sample region among those defined by the Delaunay
tessellation, and proceeds by searching for the next largest non-overlapping
sphere, until the available spheres are exhausted.

\begin{figure}
 \centering{\includegraphics[width=7.5cm]{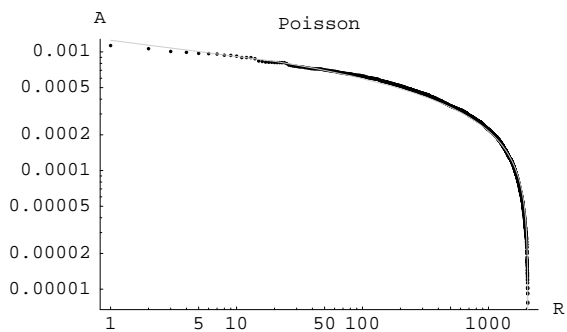}}
 \caption{Rank-ordering of the circular voids in a random set of
 $10\hspace{1pt}000$ points ($A$ is the void area and $R$ is the rank) 
 compared with the predicted law (gray line).}
\label{P-voids}
\end{figure}

This void-finder is applicable to any sample, fractal or not.  In samples of
uniform distributions, we can test the law studied in
Sect.~\ref{Poisson-p}. This law refers to {\em all} the spherical voids in a
sample but the no-overlap condition removes many of them. However, the sample
of voids obtained under this condition is unbiased, arguably.  Therefore, it
has the same distribution as the total set of voids.  To test it, we have
generated a random set of $10\hspace{1pt}000$ points in the unit square and
then run the void-finder. Its output (in rank order) is compared in
Fig.~\ref{P-voids} with the analytical prediction. This prediction results
from the two-dimensional version of Eq.~(\ref{Pcum}), namely, $\mathbb{P}_>[A]
= (1 + nA)\, e^{-nA}$, where $A$ is the void area.  Given that the number of
voids with area equal to or larger than $A$, $N_>(A)$, is the rank $R$ of the
void with area $A$, we deduce the rank order by inverting $N_>(A)$.  The
agreement shown by Fig.~\ref{P-voids} is remarkable.

\begin{figure}
 \centering{\includegraphics[width=7.5cm]{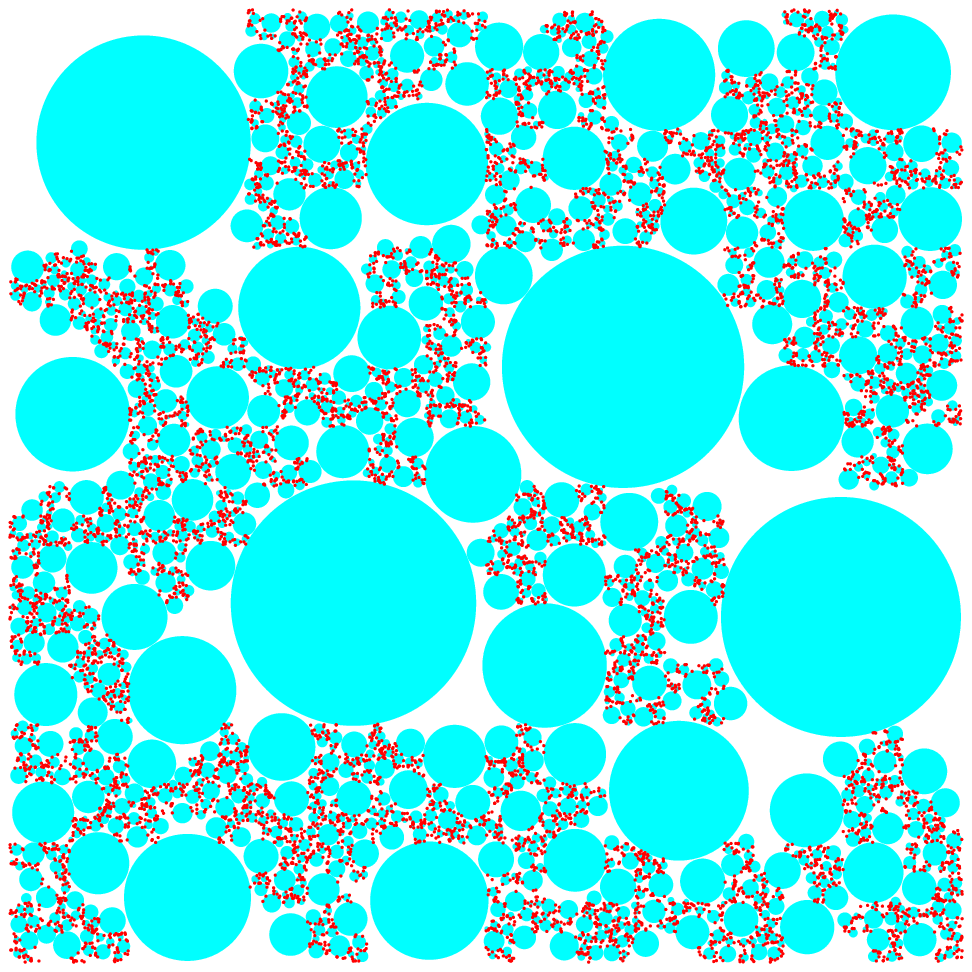}}\\[2mm]
 \centering{\includegraphics[width=7.5cm]{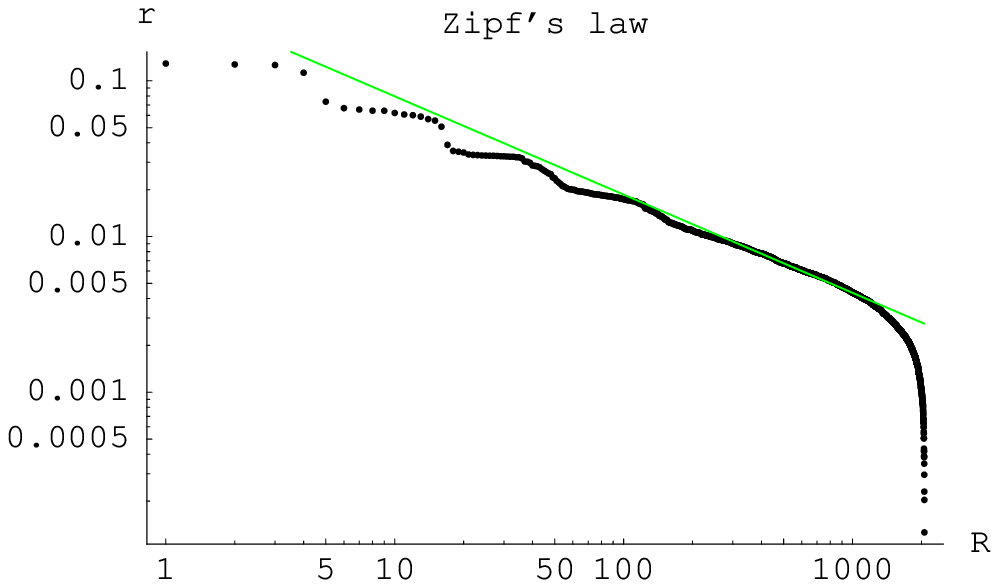}}
 \caption{(Top figure) Random Cantor-like fractal sample with
 $10\hspace{1pt}000$ points ($D=1.585$) and its corresponding voids found with
 the new algorithm (described in the text). (Bottom figure) Log-log plot of 
 the rank-ordering of the void radii, compared with the straight line 
 with slope $1/D$.}
\label{blue}
\end{figure}

To test the scaling of voids, we have applied the void-finder to several
samples of random Cantor-like fractals. We show in Fig.~\ref{blue} the results
corresponding to a random sample of $10\hspace{1pt}000$ points of a
two-dimensional random Cantor-like fractal with $D=1.585$ and compare them
with the expected Zipf law, $r \sim R^{-1/D}$ \cite{Gaite}.
Note that the found circular voids (Fig.~\ref{blue}, top) do not cover the
entire sample region (the unit square). In fact, they cover only 70\% of
it. Nonetheless, they convey well the notion of a hierarchy of voids. The
application of this void-finder to multifractal samples is explained in
Sects.~\ref{P-MFvoids}. 

\section{Geometry and scaling of voids in a monofractal}
\label{scaling}

In this section, we study the geometry of voids in monofractals and prepare
the generalization to multifractals.  Our ultimate goal is to formulate a
fractal model of voids in accord with a cosmic web structure.  We consider the
matter distribution to be continuous, unless the contrary is explicitly
stated.  The detailed treatment of continuous distributions requires some
basic notions of geometry that we now recall, for we use them in this section
and the following sections.  These notions are necessary, in particular, to
understand the intricate geometry of fractal voids.

In Euclidean space, the boundary of a region is defined as the set of points
such that any ball centered on them intersects both the region and its
complement.  A region is closed if it contains its boundary.  Normally,
fractals live in closed sets that just consist of boundary points.  A region
is open if it does not contain any boundary point. In consequence, the
complement of an open set is closed and vice versa. The union of open sets is
open.  The largest open part of a region is called its interior.  A region is
connected if it cannot be divided into two parts such that each one is
disjoint with the boundary of the other.  A region is convex if it contains
every segment with ends inside it. Finally, a set is called dense if it
intersects every open set.

Scaling of voids is natural in a strictly self-similar fractal: given that the
fractal is the union of a number of smaller similar copies of itself, every
void is reproduced on smaller scales, in an infinite hierarchy.  Mandelbrot
\cite{Mandel} expressed the similarity of voids as the diameter-number
relation $N_>(\d) \sim \d^{-D}$, where the diameter of a void is the greatest
distance between its points.  Using the rank $R(\d) = N_>(\d)$, we can write
the diameter-number relation as the Zipf law $\d(R) \sim R^{-1/D}$.  Random
fractals can only be self-similar in a statistical sense, and then voids
should adjust to the Pareto probability law $\mathbb{P}_>[\d] =
(\d_0/\d)^{D}$, with $\d_0$ being the diameter lower cutoff. This law is
indeed fulfilled by construction in one-dimensional L\'evy flights, as seen in
Sect.~\ref{nonlin-scaling}.  In higher dimensions, the geometric construction
of {\em cut-out fractals} produces a large class of fractals that are not
strictly self-similar but fulfill a form of the Zipf law
\cite{Mandel,Falc_1,I3}.  We review this construction as a model of the cosmic
foam in Sect.~\ref{cut-out}, and we also consider the scaling of voids in
general monofractals.  In Sect.~\ref{MF_cutouts}, we introduce the
generalization to non-uniform cut-out fractals, as cosmic web
models. Non-uniform cut-out fractals directly connect with multifractals,
which are the subject of Sect.~\ref{MF}.

\subsection{Cut-out fractals and cosmic foam}
\label{cut-out}

A cut-out set is obtained by removing an infinite sequence of disjoint
connected open regions from an initial region that is bounded, closed and
convex. Furthermore, for the set to be fractal, with vanishing volume, the sum
of the volumes of the removed regions must tend to the volume of the initial
region \cite{Mandel,Falc_1}.  Strictly speaking, every closed set is a
cut-out set, because its complement is open and, therefore, is the union of a
sequence of disjoint connected open regions.%
\footnote{This statement is a classic theorem of topology, proved in
Ref.~\cite{Franz}, for example.}  But it is natural to demand that proper
cut-out fractals have {\em infinite} sequences of cutouts (voids).  In one
dimension, voids are necessarily open intervals, and every cut-out fractal can
be constructed like the Cantor set, namely, by removing an infinite sequence
of open intervals.  Of course, this sequence does not need to be strictly
self-similar.  For example, the sequence of voids in a L\'evy flight is only
statistically self-similar.

In higher dimensions, a connected open region can have a very complicated
shape; for example, it can surround any number of ``islands'' and have a rough
boundary (like the shape of a cloud).  If we restrict ourselves to cut-out
fractals with convex voids and we further demand that these voids do not
degenerate to lower dimensional objects along the sequence, this sequence must
scale in a precise sense \cite{I3} (without being strictly self-similar).  The
scaling can be expressed as a particular power-law form of the rank order of
diameters: $\d(R) \asymp R^{-1/D_{\rm b}}$, in terms of the relation $\asymp$,
which means that the quotient between the related quantities is bounded above
and below.  This number-diameter relation is equivalent to a common form of
Zipf's law: the log-log plot of the rank ordering stays between two parallel
lines with slope given by the exponent ($-1/D_{\rm b}$, in the present case).
Of course, since the sequence of voids is non-degenerate, namely, the void
volume $V \asymp \d^3$, we can also express the rank-ordering as
\begin{equation}
V(R) \asymp R^{-3/D_{\rm b}}.
\label{V-R}
\end{equation}

A cut-out fractal with non-degenerate convex voids is formed by the union of
the boundaries of its voids.%
\footnote{Mandelbrot \cite{Mandel} actually studies {\em self-similar} unions
of boundaries in their own right, especially in two dimensions, under the name
of {\em sigma-loops} (sigma-loop = sum of loops).}
Thus, this type of fractals formalizes the geometry of fractal foams.
Regarding the cosmic structure, the first structures formed are actually
sheets or walls (Zeldovich's ``pancakes'') \cite{Shan-Zel}.  The sheets form
as the result of adhesive gravitational clustering: the initial under-dense
regions expand and become depleted while the walls between them concentrate
their mass, forming a foam.  In fact, foam models of the cosmic structure have
been proposed years ago.  In particular, let us mention the Voronoi foam model
by Icke and van de Weygaert \cite{Icke-Weyg}. It is natural to attribute to
the cosmic foam statistical self-similarity, which should manifest itself in
the void rank-ordering law~(\ref{V-R}).

We must examine if Eq.~(\ref{V-R}) still holds when we only consider
spherical voids, namely, the rank-ordered sequences of non-overlapping
spherical voids produced by our void-finder (e.g., the sequence of voids in
Fig.~\ref{blue}). In a cut-out fractal, each spherical void must fit in one of
its natural voids. The radius of the largest fitting sphere defines the {\em
inradius} of the void. If we denote this inradius by $r$, we have that $r
\asymp \d$, for the sequence of voids is non-degenerate. Therefore,
Eq.~(\ref{V-R}) still holds for the rank-ordered sequences of non-overlapping
spherical voids.

Finally, we consider rank-ordered sequences of voids in a fractal that is not
constructed as a cut-out set.  We can deal with this case by endowing the
fractal with a sort of ``cut-out structure'', namely, by packing the
complement of the fractal with open sets.  The geometry of the complement of a
fractal and its packing has been studied by Tricot
\cite{fractal-complement,fractal-complement_1}.  He proves a result equivalent
to Eq.~(\ref{V-R}), under several conditions on the sequence of open sets.
One important condition is the non-degeneracy of the sequence, in the sense
explained above. Another condition applies to the shape of voids and is more
general than convexity. And the condition that is crucial in the present case
is that voids must be close to the fractal, namely, the quotients of their
distances to the fractal by their diameters must be bounded. This latter
condition is certainly satisfied if the voids always touch the fractal.

\subsection{The cosmic web as a non-uniform cut-out fractal}
\label{MF_cutouts}

Let us consider a cut-out set supported on the boundaries of its voids but
with a {\em non-uniform} mass distribution on them.  We regard it as an
improved model of the cosmic fractal foam that arises from adhesive clustering:
the foam evolves with the motion of the matter in the walls towards their
intersections to form filaments, and the motion along the filaments to form
nodes, resulting in a very non-uniform distribution. This distribution of
sheets, filaments and nodes is called the cosmic web. 

For illustrating the structure of non-uniform cut-out fractals, we
construct a non-uniform fractal foam toy model that we call the {\em
Cantor-Sierpinski carpet}.  The Cantor-Sierpinski carpet is based on the
standard Sierpinski carpet and on the two-dimensional Cantor set that is the
Cartesian product of two standard Cantor sets.  The Sierpinski carpets
constitute a famous class of self-similar cut-out fractals with regular voids.
The standard triadic Sierpinski carpet is also constructed as a sort of
two-dimensional generalization of the middle third Cantor set: from an initial
square, the (open) middle sub-square of side one third is cut out, and the
iteration proceeds with the remaining eight sub-squares \cite{Mandel}. This
fractal has dimension $D = 1.89$.

\begin{figure}
 \centering{\includegraphics[width=6.5cm]{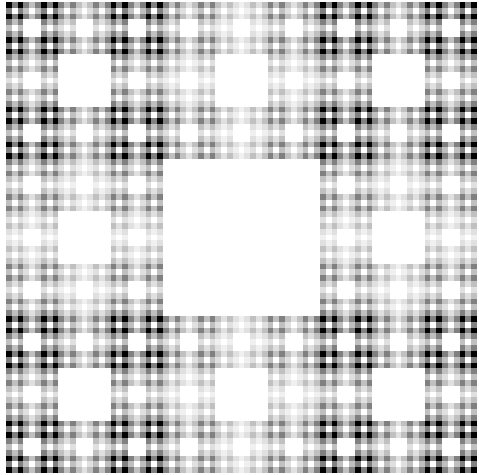}}
\caption{Two-dimensional non-uniform fractal foam: the Cantor-Sierpinski
carpet. It is a multifractal cut-out set: beside the empty voids
corresponding to the Sierpinski carpet, we can perceive very low density
regions, in contrast with the mass concentrations near the two-dimensional
Cantor dust.}
\label{C-S}
\end{figure}

We construct the Cantor-Sierpinski carpet using a modified algorithm.  The
first step still consists in cutting out the middle sub-square, an operation
that we can describe as a uniform displacement of mass from that sub-square to
the eight surrounding sub-squares. Then, to simulate the latter mass
displacement along cosmic foam walls, we further concentrate part of the mass
in the four sub-squares at the corners. Thus, the fractal generator consists
in the division of the total mass into the nine sub-squares such that the
central one receives nothing, the four sub-squares at the corners each receive
a proportion $p_1$ of the total, and the remaining four sub-squares each
receive a proportion $p_2 < p_1$, with $4 (p_1 + p_2) = 1$. The resulting
non-uniform cut-out fractal is supported on the Sierpinski carpet but has the
highest concentrations of mass in the two-dimensional Cantor set: The case
$p_1 = 1/6,\,p_2 = 1/12$ is shown in Fig.~\ref{C-S}.  Non-uniform fractal
foams are actually multifractals.

\section{Voids in a multifractal}
\label{MF}

In a multifractal, the mass in a ball of radius $r$ centered on a point $x$
decreases as a power of $r$, with a {\em local} exponent $\a(x)$ that varies
with the point \cite{Falc,Borga}.  $\a(x)$ is well defined for every point
such that the mass in any ball centered on it is non-vanishing.  The points
that fulfill this condition form the {\em support} of the distribution, which
is necessarily a closed set.%
\footnote{In general, the support of a mass distribution is defined as the
smallest closed set that contains all the mass \cite{Falc}.}
  The density at $x$ is proportional to the $\lim_{r \ra 0}r^{\a(x)-d}$, where
$d=3$ is the dimension of the ambient space.  Points with $\a(x) < d = 3$ are
{\em singular} mass concentrations (the density diverges at the point $x$). In
a multifractal model of cosmic structure, it is natural to identify these mass
concentrations with dark matter halos \cite{I,I4}.
There are also points with $\a(x) > 3$, belonging to mass depletions, which we
study next (Sect.~\ref{two-voids}).

A multifractal is formed by the union of the sets of points with given local
dimension $\a$.  The multifractal spectrum $f(\a)$ gives the dimension of the
set of points with local dimension $\a$.  It is also useful to introduce its
Legendre transform $\tau(q) = \inf_{\a}\left[q\,\a-f(\a)\right]$, which is the
scaling exponent of $q$-moments.  Hence, one derives the R\'enyi dimensions
$D(q)=\tau(q)/(q-1)$, related to information measures.  $D(0)$ is the
box-counting dimension $D_{\rm b}$ of the distribution's support and can be
identified with the maximum of $f(\a)$.  A {\em monofractal} (or {\em
unifractal}) is the particular case in which the local dimension $\a$ is
constant throughout its support and, therefore, $f(\a)=\a=D(q)$.  In other
words, a monofractal is a uniform mass distribution on a fractal support.
This uniformity implies that statistical moments fulfill the hierarchical
relation ${\bar\mu}_k \sim {{\bar\mu}_2}^{k-1}$ (Sect.~\ref{gPvoids}), whereas
in multifractals ${\bar\mu}_k \gg {{\bar\mu}_2}^{k-1}$ \cite{I4}.

Naturally, the notion of a void is more complicated in multifractals than in
monofractals.  We can actually distinguish two types of voids.

\subsection{The two types of voids in a multifractal}
\label{two-voids}

In Sect.~\ref{scaling}, we have defined a void as a connected open region that
contains no mass.  However, this definition is, in a sense, too restrictive.
Indeed, it should suffice for a point to belong to a void that the density
vanishes at it.  Thus, we can define the void region as the set of points with
vanishing density.  Since the density vanishes in open voids, this definition
is more general.  One might think that the density can only vanish, besides in
open voids, in sets of null volume such as isolated points, curves or
surfaces; but there are other possibilities.  In a multifractal, most points
can have local dimension $\a > 3$ and, therefore, vanishing density.  Unlike
the points in open voids, those points belong to the support of the
distribution and they form mass depletions inside it.  Provided that the
support of a multifractal has non-vanishing volume, its mass depletions
normally occupy most of that volume.

Therefore, we can distinguish two types of voids in multifractal
distributions: (i) open empty regions, forming the complement of the
distribution's support, and (ii) mass depletions formed by points with $\a >
3$ which belong to the distribution's support.  As a multifractal with both
types of voids, let us consider a non-uniform fractal foam, for example, the
Cantor-Sierpinski carpet in Fig.~\ref{C-S}.  The totally empty squares in this
figure are fractal voids of the first type. They follow the Zipf
law~(\ref{V-R}) with an exponent given by the box-counting dimension $D_{\rm
b}$ of the distribution's support (the Sierpinski carpet).  In this support,
there are very low density regions, which are shown in light grey in
Fig.~\ref{C-S}. In the limit of infinite iterations, the low density regions
become voids of the second type, with $\a > d = 2$ and vanishing density.

Since voids of the first type are typical of monofractals, we regard only
voids of the second type as proper multifractal voids.  Multifractals with
$D_{\rm b} < d$ have fractal support and, therefore, the voids of the second
type occupy vanishing volume.  But these voids become important when $D_{\rm
b} = d$ and there are no voids of the first type.  The case $D_{\rm b}=d$ is
already studied in Sect.~\ref{nonlin-scaling} as the case in which the
void probability function vanishes in the continuum limit, corresponding to a
non-lacunar fractal.

In this regard, we notice that the multifractal analysis of cosmological
$N$-body simulations suggests that the value of $D(0)=D_{\rm b}$ is close to
three \cite{Valda,Colom,Yepes}.  Our recent analysis of simulations of cold
dark matter \cite{I4} or of cold dark matter plus baryons \cite{I5} confirm
it.  Indeed, the supports of the distributions generated in $N$-body
simulations seem to be their entire regions of definition, that is to say, the
distributions appear non lacunar.

Let us analyze further the structure of non-lacunar multifractals.  It is easy
to realize that the geometry of proper multifractal voids is very
intricate. Note that all their points are boundary points, since these voids
have no interior.  Let us identify every connected component with an
individual void and assume that their number is countable, that is to say,
they can be arranged in a sequence.  In spite of belonging to a scaling
distribution, these voids cannot satisfy the Zipf law (\ref{V-R}), for it
would imply that the total volume of the voids $\sum_R V(R)$ diverges (since
$D_{\rm b}=3$).  On the other hand, we can have an uncountable number of
individual voids, unlike in the case of open voids.  If there is an
uncountable number of voids, every individual void can have zero volume, in
spite of their total volume being positive.  This happens, for example, in the
mass distribution produced by the one-dimensional adhesion model (ruled by the
Burgess equation with random initial conditions) \cite{V-Frisch}.%
\footnote{These properties of the voids produced by the one-dimensional
  adhesion model
follow from the properties of the complementary mass concentrations at the
shock locations: these locations form a countable and dense set.}
The cosmic web produced by the three-dimensional adhesion model
\cite{adhesion} is a complex multifractal \cite{V-Frisch,Bou-M-Parisi}, whose
geometrical features are not well understood yet.

One way to simplify the geometry of proper multifractal voids is to smooth the
density field, as we do in the next section.  In Sect.~\ref{P-MFvoids}, we
consider discrete multifractal samples and their empty spherical voids.

\subsection{Voids as under-dense regions}
\label{excursion}

Shandarin, Sheth \& Sahni \cite{Shan} and Sheth \& van de Weygaert
\cite{Sheth-vdW} define voids as under-dense regions of a continuous density
field.  The former authors further require that a void be connected.  In
general terms, under-dense regions are mass depletions, but they do not have
the specific meaning that we have given to mass depletions in multifractals.
Moreover, a multifractal does not define a regular density field, because the
density diverges in mass concentrations.  To remove these singularities, the
density field must be smoothed. For example, it can be smoothed by using a
window function or a low-pass filter of wave-numbers.  In particular, the
lognormal model employed in Sect.~\ref{correl} can be understood as a
coarse-grain approximation to a multifractal \cite{JonesCM,I4}.  A
coarse-graining procedure is also necessary to recover a continuous
distribution from a sample of it. Thus, it is used by Shandarin, Sheth \&
Sahni \cite{Shan} to find voids in finite samples without having to consider
the problem of shapes discussed in Sect.~\ref{void-f}.

Let us see how the geometry of multifractal mass depletions changes under
coarse-graining.  A suitable coarse-graining process produces a continuous
density field. If we define voids as the sets of points where the density is
{\em strictly smaller} than a given value, say the average density, then the
voids are open.%
\footnote{ They are pre-images of open sets by a continuous function and,
therefore, they are open as well.}
Thus, multifractal mass depletions become like the monofractal voids defined
in Sect.~\ref{scaling}, although these multifractal voids are not empty.  The
entire under-dense region is also formed by a sequence of connected open
regions (individual voids).  As remarked in Sect.~\ref{scaling}, connected
open region can have very complex shapes.  But their geometry is simpler than
the geometry of proper multifractal voids which we have briefly explained
in the preceding section.

\begin{figure}
 \centering{\includegraphics[width=7cm]{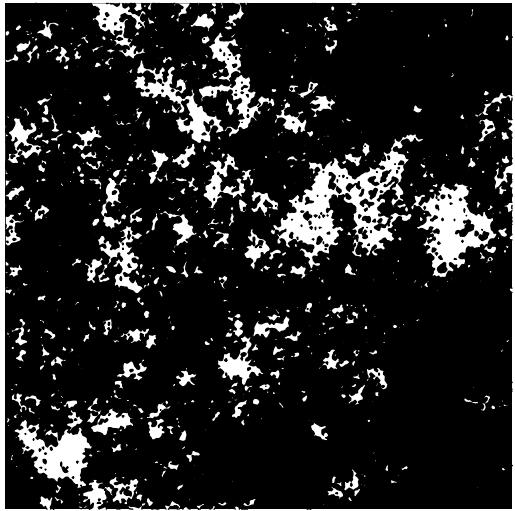}}
\caption{Voids defined by the under-dense region
of a lognormal distribution (in black).}
\label{2dFBM}
\end{figure}

For example, we have computed the iso-density level corresponding to the
average density (one) of a realization in a square of a two-dimensional
lognormal field with $\s = 1.65$ and we have plotted it in Fig.~\ref{2dFBM}.
The black region is the under-dense region, which can be decomposed into a set
of connected regions that constitute individual voids.  A good part of the
total volume belongs to the largest void, which percolates through the square.
There are smaller voids as islets inside the matter clusters (``voids in
clouds''). Unlike in a Gaussian density field, there is no symmetry between
clusters and voids: the matter clusters contain most of the mass (80\%) but
occupy a small volume (20\%).  Thus, the average density in the voids is
small, although they are not empty.

We can imagine the geometry of voids of the second type from the extrapolation
of Fig.~\ref{2dFBM} to shrinking coarse-graining lengths: more and more matter
halos pop up in the voids and more and more voids pop up in the matter
clusters. In the limit of vanishing coarse-graining length, the halos are
fully mixed with (are {\em dense} in) the voids.  This picture agrees with the
result of re-simulating with higher resolution voids in $N$-body simulations
in Ref.~\cite{Gott}.  Moreover, as the distribution becomes more singular,
voids occupy an increasing fraction of the total volume that tends to one and
contain a decreasing fraction of the total mass that tends to zero.  Despite
the large volume of this type of voids, their shape is such that no circle,
however small, can fit inside them (which is equivalent to stating that they
have no interior).

\subsection{Statistics of spherical voids in multifractal samples}
\label{P-MFvoids}

Let us show that the statistics of the spherical voids in a finite
multifractal sample provide a good indication of the type of voids. We first
consider lacunar multifractals, namely, multifractals with fractal support or,
in other words, with voids of the first type. In finite samples of these
multifractals, the size of the largest voids is hardly dependent on the
density of points $n$.  Furthermore, the sizes of voids scale.  All this
follows from the analyses in Sect.~\ref{nonlin-scaling}, Sect.~\ref{scaling}
and Sect.~\ref{two-voids}, but it is also intuitively obvious by examining
Figs.~\ref{blue} and \ref{C-S}.

\begin{figure}
 \centering{\includegraphics[width=7.5cm]{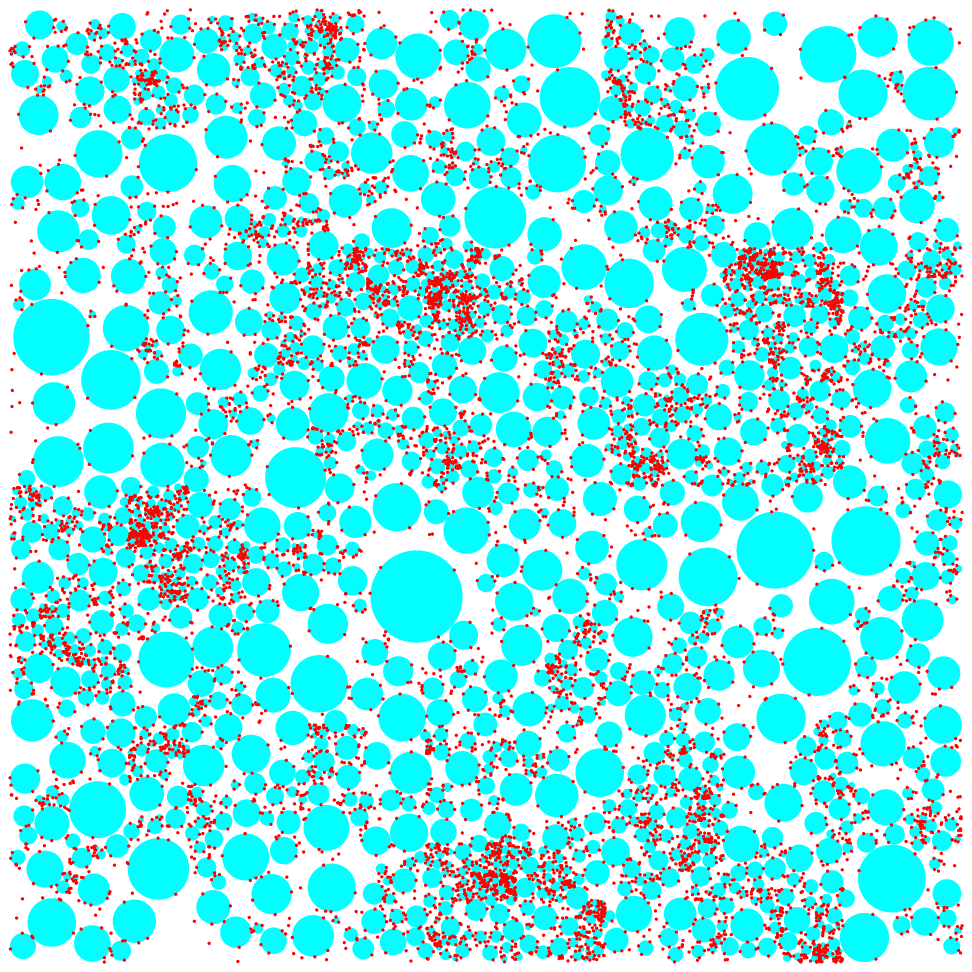}}\\[2mm]
 \centering{\includegraphics[width=7.5cm]{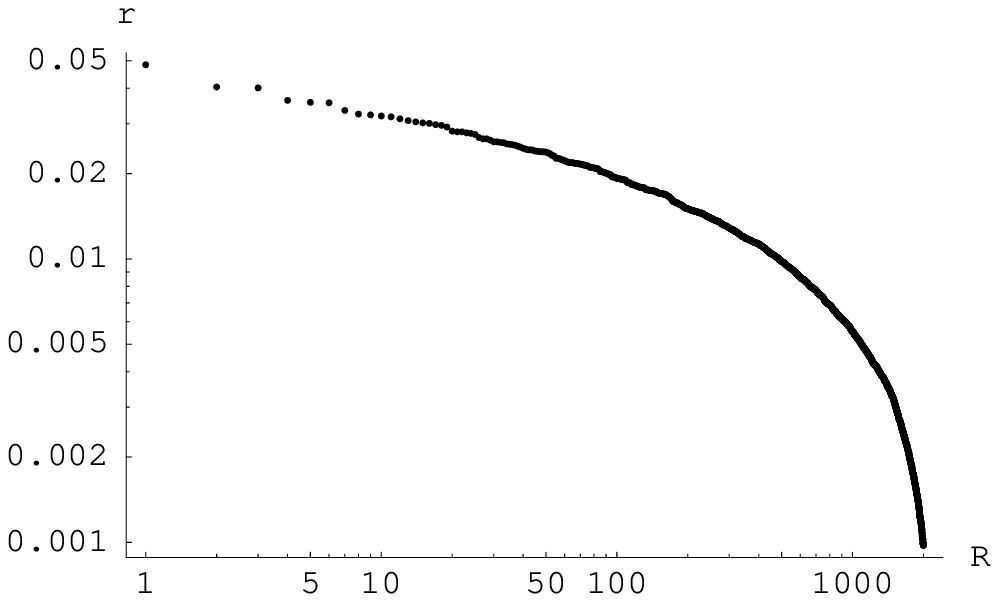}}
\caption{$10\hspace{1pt}000$ point sample of a random multifractal supported
in the full unit square, with its corresponding voids (top figure).  Log-log
plot of the rank-ordering of the void radii (bottom plot).}
\label{blue-666}
\end{figure}

For the study of the statistics of spherical voids in non-lacunar
multifractals, we use a random multinomial multifractal supported on the full
unit square.  Its correlation dimension is $D_2 = 1.526$ and its homogeneity
scale $r_0 = 1$ (the edge of the unit square).  We define this multifractal
with great precision, namely, with linear resolution $2^{24} \simeq 1.7\cdot
10^7$, and we generate a random sample of it with $10\hspace{1pt}000$ points
(the typical order of magnitude of galaxy volume-limited samples).  The
spherical voids in this sample are obtained by applying the void-finder of
Sect.~\ref{void-f}.  We find some relatively large voids (see
Fig.~\ref{blue-666}). The largest void has a radius equal to 0.0484 and an
area equal to 0.00736 (in box-size units). According to the results of
Sect.~\ref{Poisson-p}, the expected number of voids of at least that size, in
a sample of the uniform distribution with $10\hspace{1pt}000$ points and such
that $N = 73.6$, would be $10\hspace{1pt}000\,(12\pi^2/35) 73.6^2 \exp(-73.6)
\simeq 2\cdot10^{-24}$.  Therefore, the largest void is too large for a
Poisson distribution.

But the largest void is sensibly smaller than the homogeneity scale.  For this
non-lacunar fractal, we can use the estimate of the largest void size based on
the lognormal distribution that is obtained in Sect.~\ref{xtreme-c}. It is
also valid in two dimensions.  In fact, we have obtained precisely $N = 73$
for $N_{\rm t}=10\hspace{1pt}000$ points and $\s=1$. This value of $\s$ is
slightly larger than the actual value, $\s=0.82$, which can be calculated from
the point density variance at the scale of the size of the void. However,
estimates obtained from $P_0[V]$ are always lower estimates, so we conclude
that the lognormal distribution provides a reasonable model of the largest
void sizes in this multifractal.

In regard to the rank-ordering of the sequence of sizes of the spherical voids
in this multifractal sample, it does not fulfill Zipf's law, as shown in
Fig.~\ref{blue-666}. In fact, its aspect is similar to the aspect of the plot
that corresponds to a Poisson distribution (Fig.~\ref{P-voids}).

We have also tested samples of the same multifractal with larger numbers of
points.  Naturally, the largest voids are smaller in larger samples. Thus, it
becomes clear that the size of the largest void bears no relation to the
homogeneity scale. Furthermore, the estimation of that size in
Sect.~\ref{xtreme-c} based on the lognormal distribution given by
Eq.~(\ref{N1-c}), in particular, still holds in larger samples.

\section{Voids in cosmological simulations and galaxy samples}
\label{sim-gal}

\subsection{Voids in cosmological simulations}
\label{sim}

Here, we briefly study the statistics of spherical voids in data from
cosmological $N$-body simulations, taking into account the conclusions of
multifractal analyses of cosmological simulations
\cite{Valda,Colom,Yepes,I4,I5}. We have remarked above that these analyses
suggests that the box-counting dimension of the matter distribution is $D_{\rm
b}=3$, making the fractal non-lacunar. However, the measure of lacunarity
provided by $D_{\rm b}$ is rather inaccurate. Thus, it is worthwhile to study
directly the distribution of voids.

The first observation that is relevant, although somewhat trivial, is that the
spherical voids in state-of-the-art simulations, with many millions of
particles, are very small in comparison with the simulated volume and with the
scale of homogeneity as well.  This observation suggests that the matter
distribution is supported in the full simulated volume.  To confirm it, we can
measure how the size of the largest void changes with the density of points
$n$. For example, we take the redshift-zero particle distribution in the GIF2
simulation, which has a relevant scaling range, up to the homogeneity scale
$r_0 = 1/4$ (in box-size units) \cite{I4}.  We extract two random samples from
this particle distribution, with $N_{\rm t}=10\hspace{1pt}000$ and
$100\hspace{1pt}000$ points, respectively. In the first one, the largest void
has a radius equal to 0.097 and a volume equal to 0.0038 (in box-size units).
In the $100\hspace{1pt}000$ point sample, the largest void has a radius equal
to 0.059 and a volume equal to 0.00086.  This pattern of shrinking of the
largest void is consistent with a non-lacunar multifractal.

Of course, the largest void in a small sample contains particles of any larger
sample. In general, a relevant question in the process of obtaining
better samples (with larger $n$) is if and how it affects voids.  The process
cannot alter voids of the first type significantly, but its effect on proper
multifractal voids is like the effect of a shrinking coarse-graining length
described in Sect.~\ref{excursion}. Therefore, a more general and definite
proof of non-lacunarity in cosmological simulations is provided by the method
of Gottl\"ober et al \cite{Gott}, consisting in re-simulating with higher
resolution voids found in $N$-body simulations. The result of this method is
that new structure arises inside these voids in a self similar pattern, just
like a non-lacunar fractal predicts.

\subsection{Voids in galaxy samples and galaxy bias}
\label{bias}

There are many studies of the various aspects of galaxy voids. The studies
that are most relevant in our context have been carried out by Tikhonov and
Karachentsev \cite{Tikho-Kara,Tikho,Tikho2}.  They study the statistics of
voids in several galaxy samples and find evidences of scaling of galaxy voids.
As we have shown, scaling of voids is the hallmark of a lacunar fractal.
Here, we re-analyze Tikhonov's data from the 2dF survey \cite{Tikho} in regard
to the size of the largest void and, especially, to the possible scaling of
its voids.

Tikhonov \cite{Tikho} selects from the 2dF survey a volume limited sample
(VLS) with 7219 galaxies, such that the volume per galaxy is 513 Mpc$^3$
$h^{-3}$. The largest void in it corresponds to a sphere with radius 21.3 Mpc
$h^{-1}$ and volume $4.05\cdot 10^4$ Mpc$^3$ $h^{-3}$.  Thus, if the sample
belonged to a uniform distribution, the expected number of sample galaxies in
that sphere would be 79.  Then, the expected number of voids of that size,
according to Sect.~\ref{Poisson-p}, would be $7219\,(12\pi^2/35)\, 79^2\,
e^{-79} = 8\cdot 10^{-27}$, that is, absolutely negligible.  However, the
value $N=79$ is reasonable for a non-lacunar fractal, according to
Sect.~\ref{xtreme-c}.

\begin{figure}
 \centering{\includegraphics[width=7.5cm]{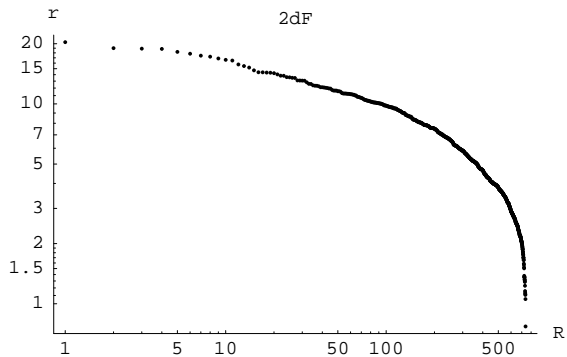}}
\caption{Rank-ordering of the voids in a 2dFGRS sample with $7\hspace{1pt}000$
galaxies ($r$ is the void radius in Mpc $h^{-1}$). This rank-ordering is
typical of non-lacunar fractals.}
\label{2dF-voids}
\end{figure}

Tikhonov \cite{Tikho} rank-orders the void sizes in the 2dF VLS and concludes
that there is a scaling range.  This scaling range is about a decade, namely,
an order of magnitude in the rank (from rank 60 to rank 600, approximately).
Tikhonov uses his own void-finder, which first fits the largest empty spheres
and then applies a merging criterion (in a similar way to El-Ad \& Piran
\cite{ElAd-Pir}).  We use our void-finder (defined in Sect.~\ref{void-f}) to
just find the list of non-overlapping spherical voids in Tikhonov's VL
sample.\footnote{Actually, we have removed a few galaxies from one boundary to
make it straight, thus making the geometrical shape of the sample rectangular
(in the angular coordinates). Furthermore, we have shifted slightly the
position of the straightened edge in order to have a round number of galaxies
in the sample, namely, $7\hspace{1pt}000$ (out of the initial
$7\hspace{1pt}219$).}  The resulting rank order is plotted in
Fig.~\ref{2dF-voids}. The range of radii is nearly the same range found by
Tikhonov but {\em no} scaling can be discerned.  The rank order shown in
Fig.~\ref{2dF-voids} is similar to the ones that we found in
Ref.~\cite{Gaite}, and they all are actually typical of non-lacunar
fractals; namely, they all are similar to the rank-ordering of Poisson voids
(Fig.~\ref{P-voids}), but they span longer ranges of sizes.

Actually, the analyses of voids that attempt to find the Zipf law have been
motivated by the assumption of a {\em lacunar} fractal distribution of
galaxies.  We must reconsider this assumption, in regard to the evidences of a
multifractal distribution of (dark) matter with $D_{\rm b} = 3$ \cite{I4,I5}
and the results of the preceding section.  Indeed, a galaxy distribution that
closely follows the full matter distribution is probably also non-lacunar.  To
actually determine the type of galaxy voids, we need to model the galaxy
biasing.

Dark matter concentrations (halos) are the natural places for galaxy
formation. These concentrations have been studied as peaks of a Gaussian
density field \cite{Kaiser}.  In contrast, our model of galaxy biasing stems
from the definition of halos as over-dense regions of a coarse-grain
multifractal density field in Sect.~\ref{excursion}.  As emphasized there,
this type of density field is very different from a Gaussian field on
nonlinear scales.  The main difference is the strong asymmetry between
over-dense and under-dense regions: the former occupy a small volume but
contain most of the mass.  Therefore, we do not need a high density threshold
to define clustered mass concentrations, but just the average density.  We can
assign a galaxy to every coarse-grain mass concentration (halo), obtaining a
clustered distribution with a minimal bias.

The average density corresponds to the local dimension $\a = 3$, which
separates mass concentrations ($\a< 3$) from mass depletions ($\a > 3$).
Naturally, we can set a higher density threshold to differentiate one highly
clustered and luminous population of galaxies (``wall galaxies'') from another
more homogeneous and less luminous population (``field galaxies'').  A natural
choice of threshold should be the density that corresponds to the local
dimension $\a_1$ of the mass concentrate \cite{I4}, because halos with smaller
densities contain relatively little mass.  Notice the analogy with the
procedure of El-Ad \& Piran \cite{ElAd-Pir} in the ``wall builder'' phase of
their void-finder.%
\footnote{Indeed, the relation between an $\a$-threshold and their procedure
  is quite close.  Since $\a(x)$ measures the
  concentration of mass around $x$, the mass in the ball of radius $r$
  centered on ${x}$ is $m[B({x},r)] \sim r^{\a(x)}.$ A discrete measure of
  $m[B(x,r)]$ is given by the number of points inside $B(x,r)$.  A threshold
  for $\a$ sets a threshold for this number, as El-Ad \& Piran do.}
With this prescription, the voids in the ``wall galaxies'' contain ``field
  galaxies''. However, the distribution of these ``field galaxies'' is not
  homogeneous and they tend to avoid the central parts of voids. The most
  homogeneous halo subpopulation consists of halos with $\a$ close to three,
  with so little gas that it may not form galaxies.  Peebles \cite{Pee2}
  discusses the nature of void objects and argues that the formation of
  observable objects is suppressed in voids, to the extent that it may
  constitute a challenge to our current structure formation paradigm.
  However, Furlanetto \& Piran \cite{Fur-Pir} use an excursion set model and
  conclude that the size of voids diminishes with the galaxy luminosities in
  reasonable agreement with observations. Our multifractal model is in accord
  with this conclusion.

\section{Summary and Conclusions}
\label{discuss}

We have developed statistical methods for the study of cosmic voids and we
have applied them to various distributions of observable objects.  The method
of choice depends crucially on the number density $n$ of objects. If the
density is sufficiently small for having a small number of objects in a volume
in which the Universe is homogeneous, then the objects are weakly correlated
and the analysis for Poisson voids in Sect.~\ref{Poisson} is relevant. Of
course, this case is the most amenable to analytical methods.  We have found
the distribution of spherical voids and the volume of the largest void, and we
have shown how to generalize these results to more general shapes of
voids. Not surprisingly, to observe a really large void, namely, with volume
$V \gg 1/n$, one needs an exponentially large sample, with a number of objects
$N_{\rm t} \sim e^{nV}$.  And this sample could only come from surveying
volumes exponentially larger than the homogeneity volume.

The conclusion is that the voids observed in galaxy surveys, for example, are
not due to Poissonian fluctuations. Indeed, galaxy samples have had for long
time number densities such that there are many galaxies in a homogeneity
volume.  The objects inside a homogeneity volume have correlations, which are
necessary for the existence of voids but make the application of analytical
methods more difficult.  For still moderate $n$, an expansion in powers of $n$
yields general results and, in particular, yields a formula for the
distribution of voids, although the computations are hardly feasible. It is
preferable to resort to significant models that are valid for any $n$. We have
focused on a hierarchical Poisson model and on the lognormal model, which
illustrate two different behaviors of the largest voids.

We define the hierarchical Poisson model by restricting the support of a
Poisson distribution to random clusters. Thus, we construct a point
distribution that saturates the general moment inequality ${\bar\mu}_k \geq
{{\bar\mu}_2}^{k-1}$. When ${\bar\mu}_2 \gg 1$, the clusters are
small and the distribution is strongly correlated, but then its large voids
are just the voids in the random distribution of clusters. These voids are
independent of the magnitude of the density of objects inside clusters.
Therefore, the distribution of the volume of the largest void is of
Fr\'echet-Pareto type, in contrast with the pure Poisson case, in which the
largest void vanishes when $n \ra \infty$ (Gumbel type).

It is natural to understand the hierarchical Poisson model as a real hierarchy
of clusters of clusters in a self-similar pattern. Thus, it becomes a fractal
model with hierarchical {\em and} scaling correlation functions. For example,
one-dimensional L\'evy flights are constructed so as to have scaling voids
that follow the Pareto distribution.  In contrast, the study of voids in
higher-dimensional fractals demands a good deal of geometry. A convenient
approach to fractal voids is the construction of cut-out fractals.  In
particular, a fractal foam is a possible model of the sheet structure of the
cosmic web.  Fractal foams have voids with simple shapes and scaling
sizes. These properties of voids can be generalized to other monofractal
distributions.

A monofractal distribution is characterized by its box-counting dimension
$D_{\rm b}$ and is essentially uniform throughout its support, like the
hierarchical Poisson model. However, the adhesive gravitational dynamics gives
rise to non-uniform distributions that contain filaments and nodes, in
addition to sheets.  This motivates us to consider non-uniform fractal foam
models and, ultimately, multifractal models of the cosmic web. A multifractal
is characterized by a spectrum of dimensions, but its box-counting dimension
still plays the most significant role in regard to voids. If $D_{\rm b} < 3$,
then the support of the multifractal has vanishing volume, namely, is itself a
fractal set. Therefore, the distribution of voids is independent of the
non-uniformity of the matter distribution inside its support. A more
interesting type of multifractal voids arises in the case that $D_{\rm b}=3$
and the support is the total volume.

Multifractals with support in a full volume have been called non-lacunar
fractals by Mandelbrot \cite{Mandel}.  Whether or not non-lacunar fractals
have voids is a subtle question. Certainly, they have no voids with empty
interior in which one can fit an empty sphere. However, they do have sets of
mass depletions with large volumes in which the density vanishes. We have
called them voids of second type. The lognormal model provides us with a good
description of the statistics of this type of voids as they appear in a
multifractal sample with number density $n$. We deduce, for example, that the
volume of the largest spherical void is much larger than in a Poisson
distribution but is still small in comparison with the homogeneity scale (its
distribution is actually of Gumbel type).  On the other hand, these voids do
not have scaling sizes.  A more geometrical picture of this type of voids is
achieved by representing them as under-dense regions of a smoothed density
field (Sect.~\ref{excursion}). Thus, second type voids become geometrically
similar to first type voids, although they necessarily contain mass. This mass
is part of self-similar structures that can be observed as the smoothing
length shrinks and the resolution increases.

Indeed, it has been observed by Gottl\"ober et al \cite{Gott} in cosmological
$N$-body simulations that a pattern of self-similar structure arises inside
voids when the resolution increases.  Further statistical evidence supports
that the voids in cosmological simulations belong to a non-lacunar fractal
(Sect.~\ref{sim}). Therefore, we can assert that dark matter voids are well
described as second type voids in a non-lacunar fractal.  

It is less certain that a similar description is appropriate for galaxy voids.
Admitting that the distribution of galaxies is fractal in the relevant range
of scales, we have examined the observational and theoretical arguments for
lacunarity.  Scaling of galaxy voids implies lacunarity.  We have not found
this scaling, neither in Ref.~\cite{Gaite} nor in Sect.~\ref{bias}, in spite
of the favorable results of Tikhonov and Karachentsev
\cite{Tikho-Kara,Tikho,Tikho2}.  Another argument for lacunarity is the
emptiness of voids.  Peebles argues that the formation of observable objects
is suppressed in voids \cite{Pee2}.  But it rather seems that the size of
voids diminishes with the galaxy luminosities \cite{Fur-Pir}. This is also
predicted by a non-lacunar multifractal model of galaxy bias in which the mass
(or luminosity) of galaxies is ruled by the local dimension as a measure of
mass concentration (Sect.~\ref{bias}). However, very weak concentrations may
not form galaxies and remain inside voids as dark matter halos. This might
produce a low lacunarity in the distribution of galaxies.

As regards the formation of voids, we have introduced multifractal models of
the adhesion-like dynamics that gives rise to the cosmic web.  However, to
obtain non-lacunar foams, we must modify slightly the construction introduced
in Sect.~\ref{MF_cutouts}; namely, we must require that the initial
under-dense regions become depleted but {\em not empty}.  This is necessary to
obtain the proper cosmic web structure. However, the construction of a
realistic model of the formation of voids along these lines is beyond the
scope of this work.

In conclusion, a good description of cosmic voids and, in general, the
structure of matter is provided by a non-lacunar multifractal, with dimension
$D_{\rm b} = 3$ (the dimension of non-fractal distributions) and without
empty voids (first-type voids).  The distribution of galaxies is
probably non-lacunar as well, but it could have a low lacunarity.  The
peculiar structure of non-lacunar fractals can be partly responsible for the
controversy about the application of fractal geometry to the distribution of
galaxies. Indeed, that structure is very different from the structure of
monofractals (or multifractals with $D_{\rm b} < 3$) that has been normally
assumed for the distribution of galaxies.

\begin{acknowledgments}
I thank Rien van de Weygaert and the other organizers of the Royal Academy
Colloquium ``Cosmic Voids'' in Amsterdam, Dec.~2006, for the invitation to
participate in it, where I conceived the idea for this research.  
I thank Anton Tikhonov for sending me his 2dF VLS data, Fernando Barbero for
calculating the asymptotic expansions in Appendix \ref{asymptotia}, and Claude
Tricot for sending me some references.
\end{acknowledgments}

\appendix

\section{The four point integral}
\label{nb-voids}

To prove Eq.~(\ref{vol-elem}) and evaluate the function $f$ therein and its
integral, we need the explicit expression of the change of Cartesian
coordinates to coordinates adapted to the sphere defined by the four points.
We first express each point as the sum $$x_i = x_\mathrm{c} + r u_i\,,$$ where
$x_\mathrm{c}$ is the center of the sphere, $r$ is its radius and, therefore,
the vectors $u_i$ are unitary; namely,
$$
u_i = (\sin \th_i \cos\f_i,\sin \th_i \sin\f_i, \cos \th_i),
$$ 
in standard spherical coordinates.  The transformation of the twelve
coordinates $\{(x_i^1,x_i^2,x_i^3)\}_{i=1}^4$ to the twelve coordinates
$\{(x_\mathrm{c}^1,x_\mathrm{c}^2,x_\mathrm{c}^3),r,\{{\bm\th}_i\}_{i=1}^4\}$,
where ${\bm\th}_i = (\th_i,\f_i)$, introduces the Jacobian matrix of the
transformation. 
For elementary reasons of symmetry, the transformation of the four point
volume element can be expressed as
$$
d^3x_1 \,d^3x_2\,d^3x_3\,d^3x_4 = 
\nonumber\\ 
d^3x_{\rm c}\,r^8 dr\,
g({\bm \th}_1,{\bm \th}_2,{\bm \th}_3,{\bm \th}_4)\,
d^2{\bm \th}_1\,d^2{\bm \th}_2\,d^2{\bm \th}_3\,d^2{\bm \th}_4,
$$
where $d^2{\bm\th} = \sin\th\, d\th\, d\f$ is the surface element on the
unit sphere and $g$ is an unknown symmetric function. Therefore, the
calculation of the Jacobian determinant is reduced to the calculation of this
function. 

The expansion of the Jacobian determinant produces $12! =
479\hspace{1pt}001\hspace{1pt}600$ terms. Although the Jacobian matrix has
many vanishing terms, the number of non-vanishing terms in the expansion of
its determinant is huge nonetheless and a direct calculation is awkward. It is
convenient to use the formalism of exterior algebra, which provides a suitable
way to organize the calculation. We begin with writing
$$ 
dx_i^a = dx_\mathrm{c}^a + u_i^a dr + r du_i^a\,, 
$$
and, then, we can calculate the one-point volume element as the wedge
product
$$
dx_i^1 \wedge dx_i^2 \wedge dx_i^3 = 
dx_\mathrm{c}^1 \wedge dx_\mathrm{c}^2 \wedge dx_\mathrm{c}^3 + \cdots
+ r^2 \,dr \wedge d^2{\bm\th}_i\,,
$$ 
where the dots stand for the eighteen remaining terms. The total number
of terms is the number of triplets of the six coordinates
$\{(x_\mathrm{c}^1,x_\mathrm{c}^2,x_\mathrm{c}^3),r,{\bm\th}_i\}$ with no
repeated coordinates, namely, $C^6_3 = 20$.

The wedge product $\wedge_{i=1}^4 (dx_i^1 \wedge dx_i^2 \wedge dx_i^3)$ could
give rise to many terms, but most terms in the one-point volume element do not
contribute to it: for example, we can deduce from the structure of the final
result that, in the volume element of index $i$, only the terms proportional
to $d^2{\bm\th}_i$ can contribute.  We note that the four-point wedge product
is oriented and one must take its absolute value for the function $g$ to be
positive and symmetric.  After a lengthy calculation, one obtains
\begin{eqnarray*}
g({\bm \th}_1,{\bm \th}_2,{\bm \th}_3,{\bm \th}_4) =\\
|C\left(\theta _3\right) C\left(\phi _2\right) S\left(\theta _1\right)
   S\left(\theta _2\right) S\left(\phi _1\right)-C\left(\theta _4\right)
   C\left(\phi _2\right) S\left(\theta _1\right) S\left(\theta _2\right)
   S\left(\phi _1\right)-\\
C\left(\theta _2\right) C\left(\phi _3\right)
   S\left(\theta _1\right) S\left(\theta _3\right) S\left(\phi
   _1\right)+C\left(\theta _4\right) C\left(\phi _3\right) S\left(\theta
   _1\right) S\left(\theta _3\right) S\left(\phi _1\right)+\\
C\left(\theta
   _2\right) C\left(\phi _4\right) S\left(\theta _1\right) S\left(\theta
   _4\right) S\left(\phi _1\right)-C\left(\theta _3\right) C\left(\phi
   _4\right) S\left(\theta _1\right) S\left(\theta _4\right) S\left(\phi
   _1\right)-\\C\left(\theta _3\right) C\left(\phi _1\right) S\left(\theta
   _1\right) S\left(\theta _2\right) S\left(\phi _2\right)+C\left(\theta
   _4\right) C\left(\phi _1\right) S\left(\theta _1\right) S\left(\theta
   _2\right) S\left(\phi _2\right)+\\C\left(\theta _1\right) C\left(\phi
   _3\right) S\left(\theta _2\right) S\left(\theta _3\right) S\left(\phi
   _2\right)-C\left(\theta _4\right) C\left(\phi _3\right) S\left(\theta
   _2\right) S\left(\theta _3\right) S\left(\phi _2\right)-\\C\left(\theta
   _1\right) C\left(\phi _4\right) S\left(\theta _2\right) S\left(\theta
   _4\right) S\left(\phi _2\right)+C\left(\theta _3\right) C\left(\phi
   _4\right) S\left(\theta _2\right) S\left(\theta _4\right) S\left(\phi
   _2\right)+\\C\left(\theta _2\right) C\left(\phi _1\right) S\left(\theta
   _1\right) S\left(\theta _3\right) S\left(\phi _3\right)-C\left(\theta
   _4\right) C\left(\phi _1\right) S\left(\theta _1\right) S\left(\theta
   _3\right) S\left(\phi _3\right)-\\C\left(\theta _1\right) C\left(\phi
   _2\right) S\left(\theta _2\right) S\left(\theta _3\right) S\left(\phi
   _3\right)+C\left(\theta _4\right) C\left(\phi _2\right) S\left(\theta
   _2\right) S\left(\theta _3\right) S\left(\phi _3\right)+\\C\left(\theta
   _1\right) C\left(\phi _4\right) S\left(\theta _3\right) S\left(\theta
   _4\right) S\left(\phi _3\right)-C\left(\theta _2\right) C\left(\phi
   _4\right) S\left(\theta _3\right) S\left(\theta _4\right) S\left(\phi
   _3\right)-\\C\left(\theta _2\right) C\left(\phi _1\right) S\left(\theta
   _1\right) S\left(\theta _4\right) S\left(\phi _4\right)+C\left(\theta
   _3\right) C\left(\phi _1\right) S\left(\theta _1\right) S\left(\theta
   _4\right) S\left(\phi _4\right)+\\C\left(\theta _1\right) C\left(\phi
   _2\right) S\left(\theta _2\right) S\left(\theta _4\right) S\left(\phi
   _4\right)-C\left(\theta _3\right) C\left(\phi _2\right) S\left(\theta
   _2\right) S\left(\theta _4\right) S\left(\phi _4\right)-\\C\left(\theta
   _1\right) C\left(\phi _3\right) S\left(\theta _3\right) S\left(\theta
   _4\right) S\left(\phi _4\right)+C\left(\theta _2\right) C\left(\phi
   _3\right) S\left(\theta _3\right) S\left(\theta _4\right) S\left(\phi
   _4\right)|\,
\phantom{.}
,
\end{eqnarray*}
where $C = \cos$ and $S = \sin$. The integral over the angles is simplified by
taking into account the permutation symmetry. It yields
$$
\int g({\bm \th}_1,{\bm \th}_2,{\bm \th}_3,{\bm \th}_4)\,
d^2{\bm \th}_1\,d^2{\bm \th}_2\,d^2{\bm \th}_3\,d^2{\bm \th}_4 =
4!\,\frac{256\,\pi^5}{105}\,.
$$

Finally, we can substitute the radius $r$ by the volume $V= 4\pi r^3/3$, using
$V^2 dV = (4\pi/3)^3 3 r^8dr$. Therefore, the function $f$ defined in
Eq.~(\ref{vol-elem}) is
$$f = \frac{3^2}{4^3 \pi^3} g,$$
and its integral is 
$$
\int f({\bm \th}_1,{\bm \th}_2,{\bm \th}_3,{\bm \th}_4)\,
d^2{\bm \th}_1\,d^2{\bm \th}_2\,d^2{\bm \th}_3\,d^2{\bm \th}_4 =
4!\,\frac{12\,\pi^2}{35}\,.
$$

\section{Asymptotic expansions of the lognormal \protect{$P_0[V]$}}
\label{asymptotia}

Here we obtain the asymptotic expansions as $\s \ra\infty$ or $N \ra\infty$ of
the integral in Eq.~(\ref{vpf-logn}), namely,
$$
\int\limits_0^\infty 
\exp\left[-\frac{(\ln r+\s^2/2)^2}{2\s^2}- N r\right]
\frac{dr}{r}\,.
$$ An inspection shows that this integral is awkward for the standard methods
of asymptotic expansion of Laplace integrals, based on Watson's lemma and its
generalizations \cite{math}. Thus, we first transform the integral, using the
following identity:
\begin{eqnarray*}
\exp\left[-\frac{(\ln r+\s^2/2)^2}{2\s^2}\right] = \\
\frac{\s}{\sqrt{2\pi}}
\int\limits_{-\infty}^\infty dz \,\exp\left[-\frac{\s^2 z^2}{2} +
i\left(\ln r + \frac{\s^2}{2} \right) z \right]
\end{eqnarray*}
Thus,
\begin{eqnarray}
\int\limits_0^\infty \exp\left[-\frac{(\ln r+\s^2/2)^2}{2\s^2}- N r\right]
\frac{dr}{r} &=& \nonumber\\
\frac{\s}{\sqrt{2\pi}}
\int\limits_{-\infty}^\infty dz\,
\exp\left[-\frac{\s^2 z^2}{2} + i\frac{\s^2}{2} z \right]
\int\limits_0^\infty \frac{dr}{r}\, e^{-Nr} r^{iz} &=& 
\nonumber\\
\frac{\s}{\sqrt{2\pi}}
\int\limits_{-\infty}^\infty dz\,
\exp\left[-\frac{\s^2 z^2}{2} + i\frac{\s^2}{2} z \right]
N^{-iz}\,\Gamma(iz),
\label{I2}
\end{eqnarray}
where the change of integration order is valid if the integrals are
convergent. For the integral over $r$ to be convergent, we need that
$\mathop{\rm Re}(iz) = -\mathop{\rm Im}(z) > 0$.

To obtain the asymptotic expansion as $\s \ra\infty$, it is convenient to
choose $\mathop{\rm Im}(z) = 1/2$ in the integral (\ref{I2}), namely, to
define $z = \z + i/2$ with $\z$ real, for it removes the linear term in the
exponential. In translating the integral (\ref{I2}) from $\mathop{\rm Im}(z) <
0$ to $\mathop{\rm Im}(z) = 1/2$, we pick up the residue at $z=0$ of the
integrand (from the Gamma function):
\begin{eqnarray*}
\int\limits_{-i-\infty}^{-i+\infty} dz\,
\exp\left[-\frac{\s^2 z^2}{2} + i\frac{\s^2}{2} z \right]
N^{-iz}\,\Gamma(iz) = \\
2\pi +
e^{-\s^2/8} N^{1/2}
\int\limits_{-\infty}^\infty d\z\,
\exp\left[-\frac{\s^2 \z^2}{2} \right] 
N^{-i\z}\,\Gamma\left(i\z-\frac{1}{2}\right)
\end{eqnarray*}
This integral is suitable for standard methods of asymptotic expansion of
general Laplace integrals \cite{math}. They yield
$$
\int\limits_{-\infty}^\infty d\z\,
\exp\left[-\frac{\s^2 \z^2}{2} \right] 
N^{-i\z}\,\Gamma\left(i\z-\frac{1}{2}\right) =
\frac{\sqrt{2\pi}}{\s} \left( f(0) + \frac{1}{2\s^2} f''(0) + \cdots
\right),  
$$
where $f(\z) = N^{-i\z}\,\Gamma\left(i\z-1/2\right)$. For the first term, we
only need $f(0) = \Gamma(-1/2) = -2\sqrt{\pi}$.

To obtain the asymptotic expansion as $N \ra\infty$, 
we write the integral (\ref{I2}) as
$$
\int\limits_{-i-\infty}^{-i+\infty} dz\,
\exp\left[-\frac{\s^2 z^2}{2} + \left(\frac{\s^2}{2}-\ln N\right) iz \right]
\Gamma(iz)\,,
$$
and find the stationary point of the exponent, which occurs at
$$
z= \left(\frac{1}{2} - \frac{\ln N}{\s^2}\right)i.
$$ 
Let us call $\l = \ln N/\s^2 - 1/2$ and make the change of variable $z = \z -
i\l $ with $\z$ real. Thus, $\mathop{\rm Im}(z) = -\l < 0$ if $\l > 0$, that
is, if $\ln N > \s^2/2$, which is fulfilled in the $N \ra\infty$ asymptotics.
Therefore,
\begin{eqnarray*}
\int\limits_{-i-\infty}^{-i+\infty} dz\,
\exp\left[-\frac{\s^2 z^2}{2} + \left(\frac{\s^2}{2}-\ln N\right) iz \right]
\Gamma(iz) = 
\\
e^{-\s^2 \l^2/2}
\int\limits_{-\infty}^\infty d\z\,
e^{-\s^2 \z^2/2} \,\Gamma(\l+i\z)
\end{eqnarray*}
The asymptotic expansion as $\l \ra\infty$ of the last integral is
straightforward, for it requires only the standard asymptotic expansion of the
Gamma function.  After some lengthy algebra, it yields
$$
\int\limits_{-\infty}^\infty d\z\, e^{-\s^2 \z^2/2} \,\Gamma(\l+i\z) = 
\Gamma(\l) \frac{\sqrt{2\pi}}{\s} \exp\left[-\frac{(\ln \l)^2}{2\s^2} \right]
\left[1 + 
{\Mfunction{O}\left(\frac{\ln^2\!\l}{\l}\right)}\right].
$$
Hence, 
\begin{eqnarray*}
\int\limits_{-i-\infty}^{-i+\infty} dz\,
\exp\left[-\frac{\s^2 z^2}{2} + \left(\frac{\s^2}{2}-\ln N\right) iz \right]
\Gamma(iz) \approx
\\
\frac{2\pi\s}{\ln N} \exp \left\{-\frac{\ln^2\!N}{2\sigma ^2}
+\frac{\ln N}{2\sigma ^2}
\left[2 \ln \left(\frac{\ln N}{\sigma ^2}\right)+\sigma ^2-2 \right]
-\frac{1}{2\sigma ^2} \ln ^2\left(\frac{\ln N}{\sigma ^2}\right)
- \frac{\sigma ^2}{8} \right\}.
\end{eqnarray*}

\section{The probability of voids in a L\'evy flight}
\label{L-flight}

L\'evy flights are analogous to Brownian random walks, but they are based on
the L\'evy stable distributions rather than on the normal distribution. Unlike
this distribution, the L\'evy stable distributions do not have finite variance
and, in particular, the L\'evy distributions that we consider here do not have
finite mean. These distributions are defined for positive $x$ only and they
have a probability function $P(x)$ with power-law asymptotic form $C/x^{D+1}$,
where $C > 0$ and $0< D < 1$.  Every L\'evy distribution is characterised by
these two parameters and we can substitute the amplitude $C$ by a scaling
parameter $t$ such that the probability density $P_t(x) = t^{-1/D}
P_1(t^{-1/D}x)$.  The explicit expression of $P_t(x)$ is only known for a few
values of $D$, chiefly, $D=1/2$, with $P_t(x) = \left(t/\sqrt{\pi}\right)
\exp(-t^2/x)/x^{3/2}$.
The cumulative probability of a L\'evy distribution is just a function of
$t^{-1/D} x$, namely, $P_>(t^{-1/D}x)$, with the parameter $D$ only, due to
the scaling property of $P_t(x)$.  In a L\'evy flight $x(t)$ with dimension
$D$, the increment $x(t)-x(s)$ has a L\'evy distribution with parameters $D$
and $t-s$.

Brownian random walks are continuous but L\'evy flights have jumps (hence
their name). Mandelbrot \cite{Mandel} studies L\'evy flight jumps and
discusses its power-law distribution, namely, $\mathbb{P}_>[L] \propto
L^{-D}$. He points out that $\mathbb{P}_>[L]$ is ill-defined as $L \ra 0$ and
explains that it must be understood as the cumulative conditional probability
$\mathbb{P}[L >l | L > \l] = (\l/l)^{D}$. However, he does not include a
proof of this very plausible equation. We include here a simple proof and a
brief discussion, for the sake of completeness.

In general, we have that
\begin{equation}
\mathbb{P}[L >l | L > \l] = \frac{\mathbb{P}_>(l)}{\mathbb{P}_>(\l)},
\label{PLL}
\end{equation}
where $l>\l$ and $\mathbb{P}_>$ is the cumulative probability of a jump in the
L\'evy flight at an arbitrary time $s$ (we can take $s=0$).  Therefore,
$\mathbb{P}_>$ is the cumulative probability $P_>$ of a L\'evy distribution
with parameters $D$ and $t$ in the limit $t \ra 0$. Naturally, $\lim_{t\ra
0}P_>(t^{-1/D}x) = 0$, but the quotient in Eq.~(\ref{PLL}) can be finite and
non-vanishing. We can evaluate the limit with l'H\^opital's rule:
\begin{eqnarray}
\mathbb{P}[L >l | L > \l] = \lim_{t\ra
0}\frac{P_>(t^{-1/D}l)}{P_>(t^{-1/D}\l)} = \lim_{t\ra
0}\frac{dP_>(t^{-1/D}l)/dt}{dP_>(t^{-1/D}\l)/dt} =
\nonumber\\
\lim_{t\ra
  0}\frac{P(t^{-1/D}l)\,t^{-1/D-1}l}{P(t^{-1/D}\l)\,t^{-1/D-1}\l} 
= \lim_{t\ra 0}\frac{(t^{-1/D}l)^{-D-1}t^{-1/D-1}l}
{(t^{-1/D}\l)^{-D-1}t^{-1/D-1}\l}
= \frac{\l^D}{l^D}.
\end{eqnarray}
Note that we need, in this calculation, only the power-law asymptotic behavior
of the L\'evy distributions, regardless of the values of $\l$ or $l$. 

From the cumulative conditional probability, we can obtain the 
conditional probability density:
\begin{equation}
\mathbb{P}[L| L > \l] = 
\left.-\frac{d}{dl}\mathbb{P}[L >l | L > \l]\right|_{l=L} 
= \frac{D\l^D}{L^{D+1}}. 
\label{P0LL}
\end{equation}
This is the quantity used in Sect.~\ref{nonlin-scaling} (with the subscript
0). The found distribution appears in various contexts and 
is often called the Pareto distribution of parameter $\l$.

It is worthwhile to remark on the fact that the unconditioned probability of
jumps $\mathbb{P}_>(l) = \lim_{t\ra 0}P_>(t^{-1/D}x) = 0$
vanishes. The reason is technical. Let us assume that the L\'evy flight $x(t)$
is defined in the interval $0 \leq t \leq 1$.  The times at which the jumps
take place form a set that is dense in $[0,1]$ (because of its statistical
self-similarity) but that is countable and, therefore, has null measure in
$[0,1]$.

The probability of voids given by the Pareto law, Eq.~(\ref{P0LL}), suggests a
simple way of generating a discrete L\'evy flight $x(t)$, with $t =
0,1,2,\ldots$, say. We can take $x(0)=0$ and successively add to it
independent values of a random variable with the Pareto distribution. Thus,
the distribution of voids is exact, namely, it coincides with the distribution
in a continuous L\'evy flight, under the condition that every void is longer
than $\l$. However, the distribution of points does not reproduce exactly the
corresponding L\'evy distribution, but it approaches it in the long
run. Indeed, the discrete L\'evy flight approaches the continuous L\'evy
flight as the number of points grows, after a rescaling of $t$. This way of
generating L\'evy flights, mentioned in Sect.~\ref{nonlin-scaling}, is very
convenient because it does not directly involve L\'evy distributions, which
are not available in analytic form (except in a few cases).

\end{document}